\documentstyle[12pt,epsfig]{article}
\newcommand{\lbfig}[1]{\refstepcounter{fig} \label{#1} }
\newcounter{fig}

\newlength{\overeqskip}
\newlength{\undereqskip}
\newcommand{\nc}{\newcommand}
\nc{\be}{\begin{equation}}
\nc{\ee}{\end{equation}}
\nc{\bea}{\begin{eqnarray}}
\nc{\eea}{\end{eqnarray}}
\nc{\bi}[1]{\bibitem{#1}}
\nc{\lsim}{\mbox{\raisebox{-.6ex}{~$\stackrel{<}{\sim}$~}}}
\nc{\gsim}{\mbox{\raisebox{-.6ex}{~$\stackrel{>}{\sim}$~}}}
\nc{\nn}{\nonumber}
\begin{document}
%
%
\begin{titlepage}
\pagestyle{empty}
\baselineskip=16pt
\rightline{HD-THEP-01-27}
\rightline{June 2001}
\baselineskip=21pt
\begin{center}
     {\Large {\bf Cosmological magnetic fields from  photon coupling\\
 to fermions and bosons in inflation }}
\end{center}
\medskip
\begin{center}
    {\Large Tomislav Prokopec}\\

{\it Institute for Theoretical Physics, Heidelberg University,}\\
{\it Philosophenweg 16, D-69120 Heidelberg, Germany}\\


\end{center}

\baselineskip=21pt
\vskip 0.5 in

\centerline{ {\bf Abstract} }
\baselineskip=18pt
\vskip 0.5truecm
\noindent
 
 We consider several gauge invariant higher dimensional operators that 
couple gravity, gauge fields and scalar or fermionic fields and thus 
break conformal invariance. In particular, we consider terms that break
conformal invariance by the photon coupling to heavy and light
fermions. While the coupling to heavy fermions typically do not 
induce significant magnetic fields, the coupling to light fermions 
may produce observable magnetic fields when there are
a few hundred light fermions. Next we consider Planck scale 
modifications of the kinetic gauge terms of the form
$f(\phi) F_{\mu\nu}F^{\mu\nu}$ 
and $h(\psi) F_{\mu\nu}\tilde F^{\mu\nu}$, where 
$f$ and $h$ are functions of scalar and pseudoscalar fields
$\phi$, $\psi$, and $F_{\mu\nu}$,
$\tilde F^{\mu\nu}$
and the gauge field strength and its dual, respectively. For a 
suitable choice of $f$ sufficiently strong magnetic fields may be
produced in inflation to be potentially observable today. 
The pseudoscalar coupling may lead to birefringence in inflation,
but no observable magnetic field
amplification. Finally, we show that the photon coupling to
metric perturbations produces by far too weak fields to be of
cosmological interest.

\vskip 1in

\rule{6cm}{.2mm}\newline
\noindent{\footnotesize
T.Prokopec@ThPhys.Uni-Heidelberg.DE}
\end{titlepage}

\baselineskip=20pt

%
%
\section{Introduction}

 Inflation~\cite{Guth1Linde1Albrecht-Steinhardt1} is at this moment the only
paradigm that offers explanation of structure formation consistent with
current microwave background~\cite{boomerang-maxima,MukhanovChibisov-Hawking}
and large structure observations, as well as the observed homogeneity and 
isotropy of the Universe on large scales and the flatness, horizon and
monopole problems. In this paper we consider several mechanisms in which
conformal invariance of the gauge fields in inflation is broken 
and discuss under what conditions inflation can produce magnetic
fields on cosmological scales. These fields may offer an explanation
for the seeds of the micro-gauss magnetic fields observed in spiral arms
of galaxies today~\cite{Kronberg}, but also for the intergalactic
magnetic fields that might influence the cosmological microwave 
background radiation (CMBR) and early structure formation. 
At the moment there is only an upper bound on cosmological magnetic 
fields of about $10^{-9}$~Gauss based on the current CMBR
measurements~\cite{bfs}. In this paper we argue that one {\it can} 
obtain observable cosmological magnetic fields from inflation. 
Hence, magnetic fields should 
be included into the analyses of the upcoming high precision CMBR
measurements, in particular in the measurements that include
polarization. To support this point of view we note that there is 
now evidence for intergalactic micro-Gauss magnetic 
field~\cite{ClarkeKronbergBohringer}, and for magnetic fields
of strength $10^{-9}$~Gauss correlated on $1$~Mpc~\cite{Tinyakov:2001nr}
from statistical correlations between the positions of high-energy
cosmic rays and BL Lacertae sources (quasars with jets).

 It is well know that quantum loop corrections to general relativity
induce corrections to Einstein's gravity~\cite{DrummondHathrell}
in curved space-time backgrounds.
Based on this observation Turner and Widrow \cite{TurnerWidrow} have argued
that breakdown of conformal invariance in inflation may be responsible 
for large scale magnetic fields that are potentially observable. In particular
they considered (i) the gravitational coupling ${\cal R}A^\mu A_\mu$, 
${\cal R}^{\mu\nu}A_\mu A_\nu$, that induce a photon mass at the price of
breaking gauge invariance, (ii) gauge invariant dimension six terms 
${\cal R}^{\mu\nu\rho\sigma}F_{\mu\nu}F_{\rho\sigma}/m^2$,
${\cal R}^{\mu\nu}g^{\rho\sigma}F_{\mu\rho}F_{\nu\sigma}/m^2$, 
${\cal R}g^{\mu\nu}g^{\rho\sigma}F_{\mu\nu}F_{\rho\sigma}/m^2$,  
and (iii) couplings of the photon field to other fields (scalars and axions).
Here ${\cal R}^{\mu\nu\rho\sigma},{\cal R}^{\mu\nu}, {\cal R}$
denote the curvature tensors and scalar, respectively, 
$F_{\mu\nu}=\partial_\mu A_\nu - \partial_\nu A_\mu$ denotes the field 
strength, and $m$ is a mass parameter. The result of their analysis is that 
Type (i) models may lead to strong magnetic fields today, because
the photon couples directly to the curvature, which is large in inflation,
${\cal R}\sim H^2$, and sufficiently small today so that it does not
conflict with the current observational bounds on the photon mass. 
The analyses of Type (ii) and (iii) models were inconclusive.

In Ref.~\cite{Dolgov} Dolgov has shown that the trace anomaly due to the
coupling of gauge fields to (massless) fermions may lead to 
gauge field amplification in inflation. Ratra~\cite{Ratra:1992bn}
has considered the effect of the inflaton coupling to gauge fields
of the form 
$e^{-\lambda\phi/M_P}g^{\mu\rho}g^{\nu\sigma}F_{\mu\nu}F_{\rho\sigma}$
and showed that one can obtain large scale magnetic fields 
in extended inflation. This idea was then reconsidered in the context of 
string cosmology by Gasperini, Giovannini and
Veneziano~\cite{Gasperini:1995dh} and by Lemoine and 
Lemoine~\cite{Lemoine:1995vj}, where the role of the inflaton is taken by
the dilaton field. Garretson, Field and 
Carroll~\cite{GarretsonFieldCarroll:1992} have considered 
coupling of the electromagnetic field to a pseudoscalar boson;
their conclusion was that it is difficult to obtain large scale magnetic
fields in this mechanism. 
Mazzitelli and Spedalieri~\cite{MazzitelliSpedalieri:1995} have
reconsidered {\it Type} ({\it ii\/}) models
of Ref.~\cite{TurnerWidrow} in the light of the Schwinger-DeWitt
expansion~\cite{DeWitt:1965} for a charged scalar field and argued 
that one cannot obtain large magnetic fields from these interactions. 
Calzetta, Kandus and Mazzitelli~\cite{Calzetta:1998ku} have considered
the charged scalar current in scalar electrodynamics as a source
for large scale magnetic fields; their results have been contested
by Giovannini and Shaposhnikov~\cite{Giovannini:2000dj}. Magnetic
field production from the dynamics of extra dimensions has been
considered in~\cite{Giovannini}. Finally, Maroto~\cite{Maroto:2000zu}
has recently suggested that coupling of gauge fields to metric
perturbations may lead to significant gauge field amplification. 

In Ref.~\cite{DavisDimopoulosProkopecTornkvist} 
we have shown that the {\it backreaction} of superhorizon scalar fields
breaks conformal invariance of gauge fields in inflation, leading to
a gauge field generation mechanism. Alternatively, the photon may acquire
a mass in inflation by a Higgs 
mechanism~\cite{DavisDimopoulosProkopecTornkvist-II}, when a charged scalar
field gets an expectation value driven by a negative coupling term, 
$V_{\rm neg}\sim -g s^2\Phi^\dagger\Phi$ to the inflaton field $s$.
When applied to electromagnetism, this then results in magnetic field
generation with the spectrum $B_{\ell}\propto \ell^{-1}$, where $\ell$ is 
the correlation length. The resulting field is sufficiently strong 
\cite{DavisDimopoulosProkopecTornkvist,DavisLilleyTornkvist,dynamo} 
to seed the galactic dynamo mechanism in flat universes with dark energy 
component. The galactic dynamo may be the mechanism that explains
the micro-Gauss magnetic fields observed in many 
galaxies today~\cite{Kronberg}. An alternative is the 
Biermann battery~\cite{Biermann:1950} which may be operative on megaparsec
scales at galaxy formation~\cite{KulsrudCenOstriker}. 

Magnetic fields may also be generated 
at cosmological phase transitions~\cite{MagneticFieldsPhaseTransitions},
as a consequence of hypermagnetic field amplification in presence of 
the right handed electrons through the Abelian triangle 
anomaly~\cite{Joyce:1997uy,Giovannini:1998gp},
and in presence of a pseudoscalar field condensate~\cite{Brustein:1999rk}.
These causal mechanisms create magnetic fields which
are correlated typically on too small scales to be of relevance for
galactic and super-galactic magnetic fields today~\cite{Tornkvist:2000ay}. 
For a review of other models for cosmological magnetic field generation
see~\cite{GrassoRubinstein}.

 In this paper we focus on mechanisms for magnetic field generation 
from terms induced by gauge field couplings to other quantum fields
in presence of gravity. We restrict ourselves to terms that 
break conformal invariance in gauge invariant manner. The paper is organized 
as follows. In section~\ref{Photon coupling to gravity with heavy fermions}
we consider the evolution of gauge fields is presence of heavy fermions
in a curved space-time background. We make use of the
local Drummond-Hathrell action~\cite{DrummondHathrell} and find that
magnetic field amplification in inflation is typically very small,
unless the number of massive fermions $N_F$ is very large,
that is $N_F\gsim 10^5$. Next, in 
section~\ref{Photon coupling to gravity with light fermions} we 
consider the anomalous gauge field coupling to light
fermions~\cite{Dolgov} and find a potentially observable 
amplification of gauge fields in inflation, provided the number of
light fermions is of the order a few hundred. 
In section~\ref{Photon coupling to scalar and pseudoscalar fields} we
then consider how modification of the gauge kinetic terms
$-\frac 14 F_{\mu\nu}F^{\mu\nu}$
and $-\frac 14 F_{\mu\nu}\tilde F^{\mu\nu}$ through the coupling to
scalar and pseudoscalar fields may affect the dynamics
of gauge fields in inflation. We find that the scalar coupling may
result in significant amplification of gauge fields~\cite{Ratra:1992bn}.
On the other hand, the coupling to pseudoscalar fields leads
to an interesting example of birefringence, but results in no observable
amplification of magnetic fields. We then in
section~\ref{Photon coupling to scalar cosmological perturbations}
show that gauge field coupling to scalar cosmological
perturbations~\cite{Maroto:2000zu}
can lead only to a very minute amplification of gauge fields in inflation.
In section~\ref{Photon at preheating} we briefly discuss
preheating, and section~\ref{Discussion and concluding remarks}
contains conclusions.

\section{Photon coupling to gravity with heavy fermions}
\label{Photon coupling to gravity with heavy fermions}

\subsection{Drummond-Hathrell action}
\label{Drummond-Hathrell action}

We now reconsider the effect of coupling of gauge fields to gravity,
corresponding to Type (ii) models of
Ref.~\cite{TurnerWidrow}), and argue that one cannot obtain strong magnetic 
fields from inflation. Drummond and Hathrell~\cite{DrummondHathrell} have 
shown that the scatterings of photons on gravitons in presence of massive 
fermions with a mass $M$ (shown in diagrams on figure~\ref{figure 1}) 
induce a correction to the Einstein-Hilbert action which, when computed in
the $1/M^2$ expansion, leads to the following effective action
\begin{eqnarray}
{\cal S} &=&\int d^4 x \sqrt{-g}\;\left[\frac{{\cal R}}{16\pi G}
  -\frac{1}{4} g^{\mu\rho}g^{\nu\sigma}F_{\mu\nu}F_{\rho\sigma} +{\cal L}_1
  \right]
\nonumber\\
{\cal L}_1 &=& 
- \frac{\beta_e}{4M^2}\left[
 b{\cal R} g^{\mu\rho}g^{\nu\sigma}F_{\mu\nu}F_{\rho\sigma}
  + c{\cal R}^{\mu\nu}g^{\rho\sigma}F_{\mu\rho}F_{\nu\sigma}
   + d{\cal R}^{\mu\nu\rho\sigma}F_{\mu\nu}F_{\rho\sigma}
\right]
\label{magf.1}
\end{eqnarray}
where $G= (8\pi M_P^2)^{-1}$ denotes Newton's constant,
$\beta_e=\alpha_e N_F/180\pi$, $\alpha_e=e^2/4\pi$, where $e$ is the photon
coupling constant at the scale relevant in inflation,
$N_F$ the number of fermionic species with mass $M$,
$b=-5, c=26$ and $d=-2$, ${\cal R}$ denotes the curvature scalar,
${\cal R}^{\mu\nu}$ and ${\cal R}^{\mu\nu\rho\sigma}$ the curvature tensors,
$F_{\mu\nu}=\partial_\mu A_\nu-\partial_\nu A_\mu$ the gauge field strength, 
$g={\rm det}[g_{\mu\nu}]$ the determinant of the metric $g_{\mu\nu}$,
and $M$ the fermion mass. In case when the gauge field is non-Abelian with 
an SU(N) symmetry, $N_F$ should be replaced by $N_F-11N/2$~\cite{Dolgov}.
Coupling to non-Abelian gauge fields thus reduces $\beta_e$
and, as we shall show, damps the gauge field in inflation. This is contrary
to fermionic loops, which induce amplification of gauge field. 
The action~(\ref{magf.1}) is computed in 
the $1/M^2$ expansion, and hence should be relevant for the photon field 
dynamics only when the curvature scale, which is characterized by the 
Hubble parameter $H$, is small in comparison to the mass scale,  
$|{\cal R} | \sim H^2 \ll M^2$. Moreover, the physical photon momentum
$k_{\rm phys}$ should satisfy $k_{\rm phys}^2\ll M^2 , |{\cal R} |$.  
When $|{\cal R} |\ll k_{\rm phys}^2\ll M^2$ the term 
$-(6\beta_e/M^2)g^{\mu\rho}D_\rho (g^{\nu\eta}F_{\mu\eta})
g^{\sigma\xi} D_\xi F_{\sigma\nu}$, induced by the
one-loop vacuum polarization diagram in figure~\ref{figure 2}, may become 
relevant. For large physical momenta, $k_{\rm phys}^2\gg M^2$, 
the $1/M^2$ expansion in~(\ref{magf.1}) breaks down,
and the fermions should be treated as massless.
Dolgov~\cite{Dolgov} has shown that in this case the fermion loop in 
figure~\ref{figure 2}  contributes to the stress-energy trace anomaly, which 
we consider in section~3. The problem of magnetic field growth in inflation
in terms of the Schwinger-DeWitt expansion~\cite{DeWitt:1965} for 
scalar fields has been studied in~\cite{MazzitelliSpedalieri:1995}.
\begin{figure}[htbp]
\begin{center}
\epsfig{file=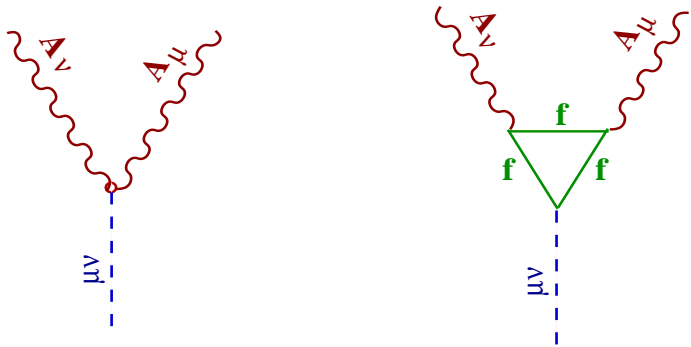, height=1.8in,width=3.6in}
\end{center}
\vskip -0.1in
\lbfig{figure 1}
\caption[fig1]{%
\small
The scattering diagrams that contribute to the Drummond-Hathrell action.
The gravitons ($G_{\mu\nu}$) are represented by the {\it dashed blue} lines, 
the photons ($A_\mu$) by the {\it wavy red} lines and the fermions ($f$) by the
{\it solid green} lines. 
}
\end{figure}
\begin{figure}[htbp]
\begin{center}
\epsfig{file=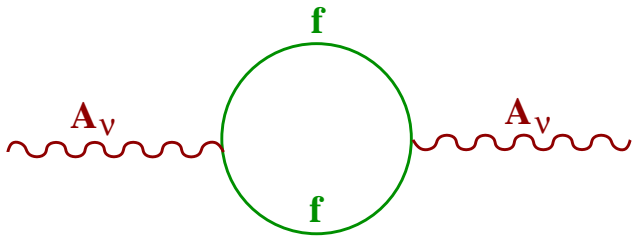, height=1.2in,width=3.3in}
\end{center}
\vskip -0.1in
\lbfig{figure 2}
\caption[fig2]{%
\small
The one-loop diagram contributing to the higher order in derivatives
dimension {\it six} operator of the Drummond-Hathrell action.
}
\end{figure}

 We shall now study the dynamics of gauge fields given by 
the Drummond-Hathrell action~(\ref{magf.1}) in conformally flat 
space-times. Recall that inflation, radiation and matter era are all
conformally flat space-times, whose metric can be parametrized as
$g_{\mu\nu}=a^2(\tau)\eta_{\mu\nu}$, where 
$\eta_{\mu\nu}={\rm diag}(1,-1,-1,-1)$ denotes the Minkowski metric,
and the scale parameter $a=a(\tau)$ is a function of conformal time $\tau$.

The equation of motion for the time-like field $A_0$ is nondynamical, and 
can be easily inferred from equations~(\ref{magf.1}) and (\ref{magf.A1}) :
\begin{equation}
 \nabla^2 A_0 - \partial_\tau \nabla\cdot \vec A = 0 \,.
\label{magf.2}
\end{equation}
which can then be used to remove $A_0$ from the dynamical $A_i$-equation.
The resulting equation is purely transverse, and identical to one
obtained in Coulomb gauge, $A_0= 0 = \nabla \cdot \vec A$. 
In Appendix~A we show that the dynamical photon field equation 
can be recast as ({\it cf.} Eq.~(\ref{magf.A6}))
\begin{equation}
\partial_\tau\left[\left(1 +  2\beta_e \frac{H^2}{M^2} \left(26 + 33w\right)
            \right)  \partial_\tau A_i^{\,T}\right]  
- \left(1 -  2\beta_e \frac{H^2}{M^2 } (10 + 3w)\right) \nabla^2 A_i^{\,T}  
      = 0  \,,\quad
\label{magf.3}
\end{equation}
where $w=p/\rho$, $\beta_e=\alpha_e N_F/180\pi$. In case when there are  
fermions with different masses, $\beta_e/M^2$ should be replaced by 
$\sum_i\beta_i/M^2_i$, where $\beta_i=\alpha_e N_{F\,i}/180\pi$.
For example, when the number of charged massive fermions is of the order 
$N_F\sim 10^3$, which is to be expected in a typical grand-unified theory, 
and taking $\alpha_e\sim 1/30$, we then have $\beta_e\sim 1/15$. 
Since typically $H/M\ll 1$, the terms that break conformal invariance 
are small. We shall now make this statement more quantitative. 

\subsection{Photon field in inflation}
\label{Photon field in inflation}

Note first that in de Sitter inflation ($w\equiv p/\rho = -1$)
Eq.~(\ref{magf.3}) becomes conformally invariant
\begin{equation}
\left(1-\frac{14\beta_eH^2}{M^2}\right)
   (\partial_\tau^2-\nabla^2)A_i^{\;T} =0 ,
\label{magf.4}
\end{equation}
which is in agreement with Ref.~\cite{DrummondHathrell}. 
Hence, in de Sitter inflation there is no photon field amplification. 
For power law inflation however conformal invariance {\it is} broken.
Consider now the homogeneous mode, for which equation~(\ref{magf.3})
is solved by
\begin{equation}
\partial_\tau A_i^{\,T}
      = C_0 \left(1 +  2\beta_e \frac{H^2}{M^2} (26 + 33w)\right)^{-1}
    \,,\quad
\label{magf.5}
\end{equation}
where $C_0=C_0(\vec x\,)$ characterises the initial electric field
in inflation. For $w < - 26/33$ the field $\partial_\tau A_i^{\,T}$ 
grows in inflation. When $2\beta_e ({H^2}{M^2}) (26 + 33w)\approx -1$
a large amplification may result. In this case adiabaticity is broken 
for superhorizon modes, leading to amplification. 

 To make this more quantitative, we now quantize the field as follows
\begin{equation}
\vec A(\tau,\vec x) = a^{-\zeta}
 \sum_{T=1,2}\int \frac{d^3k}{(2\pi)^3} \left[e^{i\vec k\cdot\vec x}
 {\cal A}^{(1)}_{\vec k}(\tau) \epsilon_{k}^T a_{\vec k}^T + 
e^{-i\vec k\cdot\vec x}
 {\cal A}^{(2)}_{\vec k}(\tau) \epsilon_{k}^T a_{\vec k}^{T\;\dagger} 
  \right] ,
\label{magf.6}
\end{equation}
where ${\cal A}^{(i)}_{\vec k}$ ($i=1,2$) denote the gauge field mode 
functions, $\epsilon_{\vec k}^T$ ($T=1,2$) are the transverse polarization
vectors, and $a_{\vec k}^T$ and $a_{\vec k}^{T\;\dagger}$ the photon field
annihilation and creation operators, respectively, which satisfy
$\left[a_{\vec k}^T,a_{\vec k}^{T^\prime\;\dagger}\right] 
  = \delta_{TT^\prime}\delta(\vec k-\vec k^\prime)$ and 
$\left[a_{\vec k}^T,a_{\vec k}^{T^\prime}\right] = 0 =
   \left[a_{\vec k}^{T\;\dagger},a_{\vec k}^{T^\prime\;\dagger}\right]$. 
The field rescaling $a^{-\zeta}$ in equation~(\ref{magf.6}) can be anticipated 
as a consequence of broken conformal invariance. The gauge field
quantization~(\ref{magf.6}) is analogous to the {\it conformal vacuum} 
of the scalar field theory in inflation. In the case of a scalar field in
conformal space-time the procedure is quite straightforward.
The kinetic term has a non-canonical form, $\sqrt{-g}{\cal L}_{\rm kin}(\phi) =
   a^2\eta^{\mu\nu}(\partial_\mu\phi)\partial_\nu\phi)/2$. The following
simple rescaling $\partial_\tau\phi=\partial_\tau\varphi/a$ brings it to
the canonical form, $\sqrt{-g}{\cal L}_{\rm kin}(\varphi)\rightarrow 
\eta^{\mu\nu}(\partial_\mu\varphi)(\partial_\nu\varphi)/2$. Since 
the canonical momentum is now $\pi_\varphi=\partial_\tau\varphi$, one can 
quantize $\varphi$ by making use of the standard canonical quantization,
$\left[\varphi(\tau,\vec x\,),\pi_\varphi(\tau, \vec x^{\;\prime}\,)\right]
  = i\delta(\vec x-\vec x^{\;\prime}\,)$. The Drummond-Hathrell
action~(\ref{magf.1}) on the other hand contains quite a nontrivial
modification of the kinetic gauge term. How to determine the power
$\zeta$ in equation~(\ref{magf.3}) is discussed in some detail below. 

From the Friedmann equation~(\ref{magf.A5}) and $\rho\propto a^{-3(1+w)}$, 
one can infer $\partial_\tau a/a= 2/(1+3w)\tau$,   
$a\propto \tau^{2/(1+3w)}$ (in de Sitter inflation  
 $\partial_\tau a/a= -1/\tau$ and $a= -1/H\tau$), so that
equation~(\ref{magf.3}) can be recast as 
\begin{equation}
\left(\partial_\tau^2 - (1-\delta_w)\nabla^{\;2} - \frac{2\theta_w}{\tau}
\partial_\tau \right) A_i^{\,T} = 0 .
\label{magf.7}
\end{equation}
where 
\footnote{In fact both $1-\delta_w$ and $\theta_w$ in Eq.~(\ref{magf.8})
should be multiplied by $(1-\iota_w)^{-1}$. This however generates terms 
that contribute to the effective action at higher orders
in the $1/M^2$ expansion, so that they can be consistently omitted.}
\begin{eqnarray}
\delta_w &=& {2\beta_e (10 + 3w)\frac{H^2}{M^2 }}
\nonumber\\
 \theta_w &=& 
 - \frac{3(1+w)}{(1+3w)}\; {\iota_w}
 \,,\quad\qquad 
\iota_w = -  2\beta_e \left(26 + 33w\right) \frac{H^2}{M^2}\,
\label{magf.8}
\end{eqnarray}
are slowly varying functions of conformal time. Since 
$H^2\propto \rho\propto \tau^{-6(1+w)/(1+3w)}$, this is indeed so
in slow-roll inflation, for which $w\approx -1$. When the time dependence
of the Hubble parameter is neglected, equation~(\ref{magf.7})  
can be reduced to the following Bessel equation
\begin{equation}
\left(\partial_\tau^2 + \vec k ^{\;2} - 
  \frac{\theta_w\left(1+\theta_w\right)}{\tau^2}
 \right) {\cal A}_{\vec k} = 0 ,
\label{magf.9}
\end{equation}
where the mode functions ${\cal A}_{\vec k}$ are defined in 
equation~(\ref{magf.6}), and we set $\delta_w\rightarrow 0$. 
This is legitimate for superhorizon modes, since the effect induced by 
$\delta_w$ is suppressed by $k^2$ when compared with the
$\theta_w$-contribution. In Appendix~A we also consider 
the scalar electrodynamics~\cite{MazzitelliSpedalieri:1995}
and show that for the minimal coupling $\xi=0$ and 
when $\xi<1/6$ the gauge fields get damped in inflation.
In the conformally coupled case ($\xi=1/6$) there is essentially
no effect, while in the case when $\xi>1/6$ the
gauge fields grow in inflation ({\it cf.} Eq.~(\ref{magf.A11})). 

In deriving equation~(\ref{magf.9}), the requirement 
that the coefficient of the damping term $\partial_\tau A^T_i$ vanishes
leads to the following conformal mode rescaling 
$a^{-\zeta}\propto (-\tau)^{\theta_w}$ in equation~(\ref{magf.6}), and hence 
\begin{equation}
a^{-\zeta} = a^{\frac{(1+3w)}{2}\,\theta_w}\,.
\label{magf.10}
\end{equation}
With this conformal rescaling the quantization rules for the photon field
in~(\ref{magf.6}) are canonical, that is 
$[a_{\vec k}^T, a_{\vec k^\prime}^{T^\prime\;\dagger}]
  = \delta^{TT^\prime}\delta(\vec k-\vec k^\prime)$ and the Wronskian
    ${\bf W}[{\cal A}_{\vec k}^{(1)},{\cal A}_{\vec k}^{(2)}] = i$. 

The solution to equation~(\ref{magf.9}) can be conveniently expressed in terms
of Hankel functions as follows
\begin{equation}
{\cal A}^{(j)}_{\vec k} = \frac{1}{2}\sqrt{-\pi\tau}
   \;H^{(j)}_\nu(-k\tau)\,,\quad j=1,2,\qquad
          \nu = \theta_w + \frac{1}{2} \,
\label{magf.11}
\end{equation}
which, at early times in inflation when $\tau\rightarrow -\infty$, 
reduce to the conformal vacuum solutions
\begin{equation}
{\cal A}^{(j)}_{\vec k} \stackrel{\tau\rightarrow - \infty}{\longrightarrow}  
  \frac{1}{\sqrt{2k}} e^{\mp i\left(k\tau + \pi\nu/2 + \pi/4\right)}
+ o((-k\tau)^{-1}) 
  \,,\quad j=1,2,\quad
\label{magf.12}
\end{equation}
with the standard Wronskian wave function normalization
\begin{equation}
{\bf W}\left[{\cal A}^{(1)}_{\vec k},{\cal A}^{(2)}_{\vec k}\right] \equiv
 {\cal A}^{(1)}_{\vec k}\partial_\tau{\cal A}^{(2)}_{\vec k}
 - \left(\partial_\tau {\cal A}^{(1)}_{\vec k}\right){\cal A}^{(2)}_{\vec k}
   = i .
\label{magf.13}
\end{equation}
When writing Eq.~(\ref{magf.11}) we assumed that $\theta_w>-1/2$, 
so that $\nu>0$, which holds in situations of physical interest. 

 We are primarily interested in computation of magnetic fields on cosmological
scales today, which correspond to superhorizon scales at late time in
inflation, for which $k_{\rm phys}\ll H$. For these scales $k|\tau|\ll 1$,
and one can use the small argument expansion for the Hankel functions
$H_\nu^{(1,2)}$ in~(\ref{magf.11}) to get  
\begin{equation}
{\cal A}^{(1)}_{\vec k} = {\cal A}^{(2)\;*}_{\vec k} = 
\frac{\sqrt{-\pi\tau}}{2}\left[
   -i\frac{\Gamma(\nu)}{\pi}\left(-\frac{2}{k\tau}\right)^{\nu}
+ \frac{1-i\cot\pi\nu}{\Gamma(\nu+1)}
     \left(-\frac{k\tau}{2}\right)^{\nu}
\, \right]  \left(1-o((k\tau)^2)\right),
\label{magf.14}
\end{equation}
where we assumed $k|\tau|\ll 1$, $\nu>0$ and $\nu\neq$~integer. The expansion
of Hankel functions for $\nu=$~integer can be found for example in 
Ref.~\cite{GradshteynRyzhik}.

 To model the gauge field evolution in radiation and matter eras, two 
important effects must be taken account of : (a) the energy density 
$\rho$ begins scaling away quickly with the expansion of the Universe, 
$\rho\propto 1/a^4 \propto 1/\tau^4$, and, (b) as a consequence of the
inflaton decay into charged particles, the plasma conductivity grows large. 
Usually the transition to radiation era
is fast, and typically takes less than one expansion time, while 
the inflaton decay to radiation is model dependent and may take many expansion
times, so that one should consider these two effects separately.
It is important to make this distinction because, as we shall see, 
a consequence of the inflation-radiation matching may be
the photon field growth, while a large conductivity induces 
freeze-out of the magnetic field and decay of the electric field.

\subsection{Photon field in radiation era}
\label{Photon field in radiation era}

 We shall now compute the gauge field in radiation era in the sudden
transition approximation. That means that we neglect the 
effect of conformal invariance breakdown, which may play some role at early
stages of radiation era. In addition we assume that
the effect of conductivity and preheating can be neglected. 
(For a heuristic account of the conductivity at preheating 
see Ref.~\cite{DavisDimopoulosProkopecTornkvist-II}.)
The matching conditions then become simply    
\begin{eqnarray}
{\cal A}_{\vec k}^{(1)}(\tau_i) &\equiv& {\cal A}_0 =
   \alpha_{\vec k}{\cal A}_{\vec k}^{(+)}(\tau_r)
   +  \beta_{-\vec k}^*{\cal A}_{\vec k}^{(-)}(\tau_r)
\nonumber\\
\partial_\tau {\cal A}_{\vec k}^{(1)}(\tau_i)
+ \frac{\nu-1/2}{\tau_i} {\cal A}_{\vec k}^{(1)}(\tau_i)
    &\equiv& - {\cal E}_0 = 
  \alpha_{\vec k}\partial_\tau{\cal A}_{\vec k}^{(+)}(\tau_r)
   +  \beta_{-\vec k}^*\partial_\tau{\cal A}_{\vec k}^{(-)}(\tau_r) ,
\label{magf.15}
\end{eqnarray}
where ${\cal A}^{(\pm)}_{\vec k}
=(2 k)^{-1/2}e^{\mp i k\tau}$ are 
the asymptotic (conformal) vacuum solutions in radiation era.

 Determining the conformal times at the end of inflation ($\tau=\tau_i$) and
the beginning of radiation era ($\tau=\tau_r$) requires some care. 
We start with the physical requirement, that the photon field is
in the (Minkowski) vacuum for $k_{\rm phys}\equiv k/a  \geq M$.
We are free to set $a=1$ at the conformal time 
when the vacuum changes from Minkowski to conformal, that is when the terms
in Eqs.~(\ref{magf.1}) and~(\ref{magf.3}) that break conformal 
invariance become operative.  
The modes $k=M$ then exit the horizon ($k|\tau|=1$) at $a=M/H$. This then
implies the following functional form of the scale factor in inflation 
\begin{equation}
a(\tau)= \frac{M}{H}(-M\tau)^{2/(1+3w)}\,,\qquad \tau\leq \tau_i=-\frac{1}{M}
.
\label{magf.16}
\end{equation}
The smooth matching $a(\tau_i) = a(\tau_r)$
and $\partial_\tau a(\tau_i) = \partial_\tau a(\tau_r)$ then implies 
for the scale factor in radiation 
\begin{equation}
a(\tau)= \frac{M}{H}\frac{\tau}{\tau_r}\,,\qquad 
  \tau\geq \tau_r=-\frac{1+3w}{2}\frac{1}{M} \,.
\label{magf.17}
\end{equation}

The nontrivial term $(\nu-1/2){\cal A}_{\vec k}^{(1)}(\tau_i)/\tau_i$ 
in the second matching condition~(\ref{magf.15}) comes from 
the scale prefactor in equation~(\ref{magf.6}). 
In this case the coefficients $\alpha_{\vec k}$ and  $\beta_{\vec k}$ 
correspond to the Bogoliubov coefficients because we are matching onto
the conformal vacuum of radiation era, such that the ${\vec k}$-mode
amplitude amplification can be characterized by the number of produced
particles $N_{\vec k} = \vert\beta_{\vec k}\vert^2$. 
The result for the matching coefficients is 
\begin{equation}
\alpha_{\vec k} = \frac{{\cal A}_0-i{\cal E}_0/k}
  {2{\cal A}_{\vec k}^{(+)}(\tau_r)}
\,,\qquad
 \beta_{-\vec k}^*  = 
  \frac{{\cal A}_0+i{\cal E}_0/k}{2{\cal A}_{\vec k}^{(-)}(\tau_r)}
 ,
\label{magf.18}
\end{equation}
and hence
\begin{equation}
{\cal A}_{\vec k}(\tau) = {\cal A}_0\cos k(\tau-\tau_r)
     -\frac{{\cal E}_0}{k}\sin k(\tau-\tau_r) \,.
\label{magf.19}
\end{equation}
Now taking a small-argument expansion of the Hankel functions~(\ref{magf.14})
and assuming $\nu>0$, we obtain
\begin{eqnarray}
\nonumber\\
{\cal A}_{0} &=& - \frac{i}{2} \,\Gamma(\nu) \left(\pi M\right)^{-\frac{1}{2}}
\left(\frac{2M}{k}\right)^{\nu}
\,,\qquad \nu>0
\nonumber\\
{\cal E}_{0} &=&  
\left(\pi M\right)^{1/2}
 \frac{(1-i\cot\pi\nu)}{\Gamma(\nu)}\left(\frac{k}{2M}\right)^\nu
\,,\qquad 1> \nu>0
\nonumber\\
{\cal E}_{0} &=& - i \Gamma(\nu+1) \left(\frac{M}{\pi}\right)^{1/2}
\left(\frac{k}{2M}\right)^{2-\nu}
   \,,\qquad \nu>1\,.
\label{magf.20}
\end{eqnarray}
where we made use of $\Gamma(1-z)\sin\pi z =\pi/\Gamma(z)$, such that 
these equations are valid for integer $\nu$'s as well.
To get the photon field amplitude~(\ref{magf.6}) in radiation era,
multiplication by a factor 
$a^{-\zeta} = (H/M)^{-\frac{1+3w}{2}\left(\nu-\frac{1}{2}\right)}
= (H/M)^{-\frac{1+3w}{2}\theta_w}$
is required. When expressed in terms of $\theta_w = \nu-1/2$, 
equations~(\ref{magf.20}) become
\begin{eqnarray}
\nonumber\\
{\cal A}_{0} &=& -  \frac{i}{2}\Gamma(\theta_w+1/2)
 \left(\pi M\right)^{-\frac{1}{2}}
  \left(\frac{2M}{k}\right)^{\frac{1}{2}+\theta_w}
\,,\qquad \theta_w> -\frac{1}{2}
\nonumber\\
{\cal E}_{0} &=& \left(\pi M\right)^{1/2}
 \frac{(1+i\tan\pi\theta_w)}{\Gamma(\theta_w+1/2)}
   \left(\frac{k}{2M}\right)^{\frac{1}{2}+\theta_w}
\,,\qquad  \frac{1}{2} > \theta_w> - \frac{1}{2}
\nonumber\\
{\cal E}_{0} &=& - i \Gamma(\theta_w+3/2) \left(\frac{M}{\pi}\right)^{1/2}
  \left(\frac{k}{2M}\right)^{\frac{3}{2}-\theta_w}
   \,,\qquad \theta_w>  \frac{1}{2}
\label{magf.21}
\end{eqnarray}
From these and equation~(\ref{magf.19}) we easily conclude that 
the contribution of ${\cal A}_0$ dominates, and hence in this mechanism
the electric field contribution ${\cal E}_0$ can be neglected.
For this to occur the term $(\nu-1/2){\cal A}_{\vec k}^{(1)}(\tau_i)/\tau_i$  
in equation~(\ref{magf.15}) plays an essential role, since it cancels out 
the leading contribution to ${\cal E}_0$, making thus the ${\cal E}_0$ term 
in equation~(\ref{magf.19}) subleading. 
This means that all of the amplitude growth by the
Drummond-Hathrell action is effected in inflation, and 
the magnetic field strength can be approximated by 
\begin{equation}
{\cal A}_{\vec k}(\tau) = {\cal A}_0\cos k(\tau-\tau_r)
\label{magf.22}
\end{equation}
and is independent of conductivity in radiation era.   
This is to be contrasted with the backreaction mechanism of 
Ref.~\cite{DavisDimopoulosProkopecTornkvist}, which corresponds 
to $\theta_w+1/2\rightarrow M_A^2/H^2$, where $M_A$ is the photon mass in 
inflation, such that ${\cal E}_0/k\gg {\cal A}_0$ for superhorizon modes. 

\subsection{Magnetic field spectrum}
\label{Magnetic field spectrum}

We shall now compute the magnetic field spectrum
from the Drummond-Hathrell action in inflation and discuss it's cosmological
relevance. Since the field originates from the amplified vacuum fluctuations
in inflation, it is classical and stochastic on cosmological scales, 
so that the field can be completely specified by the spectrum. By making use
of the volume averaging procedure~\cite{DavisDimopoulosProkopecTornkvist}
described in Appendix B, we find that the gauge field 
spectrum~(\ref{magf.21}) --~(\ref{magf.22}) and equations~(\ref{magf.10})
and~(\ref{magf.B7})
imply the following spectrum at the end of inflation
\begin{eqnarray}
B_{\ell,\,\theta_w} &=& \frac{2^{\theta_w-1/2}}{\pi^{1/2}}
  \Gamma\left(\theta_w+1/2\right) b_{\theta_w +\frac 12}
\left(\frac{M}{H}\right)^{\frac{3}{2}(1+w)\theta_w}
 H^2 (H\ell)^{\theta_w-2}\,
\nonumber\\
 b_{\theta_w +\frac 12}^2 &=& \frac{9\times 2^{2\theta_w-8}}{\pi^{3/2}}
\frac{\Gamma\left(2 -\theta_w\right)\Gamma\left(\theta_w\right)}
   {\Gamma\left(\theta_w + 1/2\right)\Gamma\left(\theta_w + 2\right)}
\,.
\label{magf.23}
\end{eqnarray}
%
%
%
To obtain the magnetic field today, we assume the field to be frozen
in the plasma such that $B(T_0) = B (T_H) (T_H/T_0)^2$, and 
the comoving scale $\ell_c$ scales as  $\ell_c =  (T_H/T_0)\,\ell $, where 
$T_0=2.75\;{\rm K}\equiv 2.37\times 10^{-13}\;{\rm GeV}$ and 
$T_H\sim 10^{15}$~GeV is the temperature corresponding to the end of
inflation when $H\simeq 10^{13}$~GeV. With this the {\it spectrum today} 
can be written as  
\begin{equation}
B_{\ell,\,\theta_w} = \frac{2^{\theta_w-1/2}}{\pi^{1/2}}
  \Gamma\left(\theta_w+1/2\right) b_{\theta_w +\frac 12}
\left(\frac{M}{H}\right)^{\frac{3}{2}(1+w)\theta_w}
\left(\frac{T_0}{T_H}\right)^{\theta_w}
\frac{H^{\theta_w}}{\ell_c^{\,2-\theta_w}}\,.
\label{magf.24}
\end{equation}
%
%
%
%
When the effects of turbulence are included,
a modest amplification of the amplitude on large scales may result,
provided the spectrum slope $\theta_w -2 <-1/2$ ({\it cf.} figure~\ref{spectra}
in Ref.~\cite{DavisDimopoulosProkopecTornkvist}). 
To evaluate the magnetic field strength on cosmological scales, 
the following conversions are useful :
$(1{\rm GeV})^2 \equiv 1.44\times 10^{19}$~Gauss
and $10\,{\rm kpc}\equiv 1.56\times 10^{36}\;{\rm GeV}^{-1}$.
In figure~\ref{spectra} we show the spectra obtained for various $\theta_w$ 
of physical interest. In particular, the {\it vacuum spectrum}, which 
corresponds to the conformally invariant case, is recovered when
$\theta_w=0$, and can be inferred from equation~(\ref{magf.B9})~: 
\begin{equation}
B_{\ell_c}^{\rm vac} = \frac{3}{16\pi}\; {\ell_c^{-2}}
 \approx 3\times 10^{-55} \left(\frac{10\;{\rm kpc}}{\ell_c}\right)^2
               {\rm Gauss}\,\qquad ({\rm vacuum})\,.
\label{magf.25}
\end{equation}
The {\it thermal spectrum} corresponds to $\theta_w=1/2$ in
equation~(\ref{magf.24}) and reads
\begin{equation}
B_{\ell_c,\,\frac 12} \approx 2\times 10^{-44} 
\left(\frac{M}{H}\right)^{\frac{3}{4}(1+w)}
  \left(\frac{10\;{\rm kpc}}{\ell_c}\right)^\frac 32
               {\rm Gauss}\,.
\label{magf.26}
\end{equation}
Since the lower bound on the seed field for the galactic dynamo
is about $B_{\rm seed }\geq 10^{-35}$~Gauss at a comoving scale
of $\ell_c \sim 10$~kpc
\cite{DavisDimopoulosProkopecTornkvist,DavisLilleyTornkvist}, 
both of these spectra are too weak to be of cosmological interest. 
When $\theta_w=1$ however, equation~(\ref{magf.24}) reduces to 
\begin{equation}
B_{\ell_c,\,1} \approx 2\times 10^{-33}
\left(\frac{M}{H}\right)^{\frac{3}{2}(1+w)}
\frac{10\;{\rm kpc}}{\ell_c}\,,
\label{magf.27}
\end{equation}
which is sufficiently strong to seed the galactic dynamo. 
In other words, to obtain the field of cosmological interest, 
the spectral index must be greater than about $ -1$. An exception are 
models in which the photon becomes massive dynamically in 
inflation~\cite{DavisDimopoulosProkopecTornkvist,
DavisDimopoulosProkopecTornkvist-II}. Finally, when $\theta_w=2$ the spectrum
becomes scale invariant with the field strength 
({\it cf.} Eq.~(\ref{magf.B10}))~:
\begin{equation}
B_{\ell_c}^{\rm\, sc.inv.} \approx 3\times 10^{-11}
\left(\frac{M}{H}\right)^{3(1+w)}\;{\rm Gauss}\,.
\label{magf.28}
\end{equation}
Here we have not displayed the logarithmic enhancement 
$\sim\left(-\ln 2k_0\ell\right)^{\frac 12}$ in Eq.~(\ref{magf.B10}),
where $k_0$ is the smallest momentum amplified in inflation. Note that
for $M\sim H$ the field~(\ref{magf.28}) is somewhat weaker than
$B \sim 10^{-9}$~Gauss suggested in~\cite{Tinyakov:2001nr}. Furthermore, it 
conforms with the magnetic field bound from the CMBR
considerations in~\cite{bfs}. In figure~\ref{spectra} we illustrate
all of the spectra~(\ref{magf.25}) -- (\ref{magf.28}).  
\begin{figure}[htbp]
\begin{center}
\epsfig{file=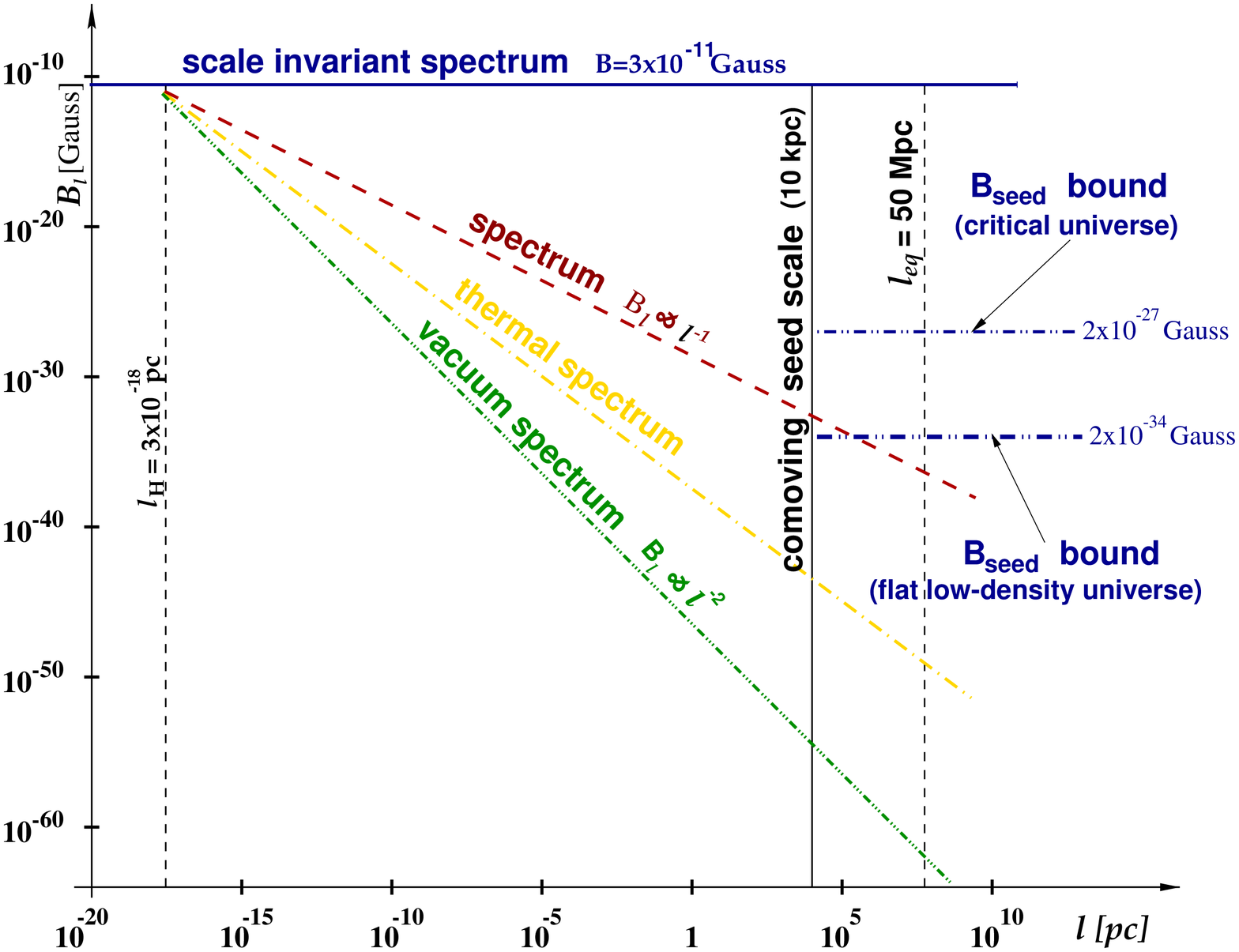, height=3.3in,width=4.9in}
\end{center}
\vskip -0.1in
\lbfig{spectra}
\caption[fig3]{%
\small Magnetic-field spectra~(\ref{magf.24}) from the Drummond-Hathrell
action as a function of the comoving scale $\ell_c$ today. The 
{\it vacuum spectrum}~(\ref{magf.25}) ($\theta_w=0$) is shown in
{\it dot-dot-dashed green}, the thermal spectrum~(\ref{magf.26}) 
($\theta_w=1/2$) in {\it dot-dashed gold}, the $\theta_w=1$
spectrum~(\ref{magf.27}) in {\it dashed red}, and the scale invariant 
spectrum~(\ref{magf.28}) ($\theta_w=2$) in {\it solid blue}, where 
we assumed that the fermion mass $M\sim H$. 
We also show the relevant dynamo bounds
$B_{\rm seed} \gsim 2\times 10^{-27}$~Gauss for a universe with critical
matter density, and $B_{\rm seed} \gsim 2\times 10^{-34}$~Gauss for a flat,
low-density universe with dark energy component.
Note that when $\theta_w\gsim 1$ the dynamo bound 
$B_{\rm seed}\gsim 2\times 10^{-34}$~Gauss at $\ell_c\sim 10$~kpc in a flat,
low-density universe is amply satisfied.}
\end{figure}

 An important question is of course whether one can obtain $\theta_w\sim 1$
from the Drummond-Hathrell action with a realistic choice of parameters.
To investigate this, we write equation~(\ref{magf.8}) as 
\begin{equation}
 N_F(\theta_w) = \frac{(1+3w)(26 + 33w)}{(1+w)}\frac{30\pi}{\alpha_e}
\frac{M^2}{H^2} \theta_w \,.
\label{magf.29}
\end{equation}
To good approximation, $\theta_w$ in equation~(\ref{magf.8}) 
maximizes when $w\simeq w_{\rm m} = -59/66$ (for which 
$-(26+33w)(1+w)=$~maximum), such that 
\begin{equation}
 N_F(\theta_w=1) \simeq  \frac{1665\pi}{\alpha_e}\,\frac{M^2}{H^2}\,,
\label{magf.30}
\end{equation}
and hence $N_F\sim  10^5$ charged fermions with a mass $M\sim H$ result in
$\theta_w\sim 1$, too high to be expected in grand-unified theories.
In this case $\iota_w\sim 30$, indicating break-down of the $1/M^2$ expansion.
Further, in this mechanism the value $\theta_w=1$ is not in any sense
favoured, or natural. In conclusion, we have shown that is it quite unlikely 
that magnetic fields can be amplified by the Drummond-Hathrell action 
in inflation to a strength of interest for cosmology. Moreover, our result
should be interpreted with caution, since large amplification is obtained in
the region of parameter space when the $1/M^2$ expansion becomes unreliable. 
In order to address this question one would have to study systematically 
higher order corrections in the $1/M^2$ Schwinger-DeWitt
expansion~\cite{DeWitt:1965}.

\section{Photon coupling to gravity with light fermions}
\label{Photon coupling to gravity with light fermions}

\subsection{Trace anomaly}
\label{Trace anomaly}

Dolgov~\cite{Dolgov} has argued that conformal invariance of 
gauge fields gets broken by the one-loop diagram shown in
figure~\ref{figure 1}, even in presence of massless fermions. 
Taking account of the local contribution only, in a conformal 
space-time with the metric $g_{\mu\nu}=a^2\eta_{\mu\nu}$,   
the following effective action 
\begin{equation}
{\cal S}_{\rm anom} = \int d^{\,4} x (-g)^{\frac{1}{2}+\frac{\kappa_e}{8}}\;
\left[  -\frac{1}{4}
   g^{\mu\nu}g^{\rho\sigma}F_{\mu\nu}F_{\rho\sigma}
  \right],
\label{magf.31}
\end{equation}
where 
\begin{equation}
\kappa_e(N_F)= \frac{2\alpha_e N_F}{3\pi}
\label{magf.32}
\end{equation}
reproduces the equation of motion in Ref.~\cite{Dolgov}.
This action manifestly breaks conformal invariance, since it is not 
invariant under the transformation $a\rightarrow \lambda a$. 
It is however invariant under the following combination of 
conformal transformation and dilatation,
$a\rightarrow \lambda a$, and $x^\mu\rightarrow \lambda^\prime x^\mu$,
with $\lambda^\prime =\lambda^{-\kappa_e/2}$.  

When non-Abelian gauge fields with an SU(N) symmetry are considered, 
the anomaly can be studied by 
$\kappa_e\rightarrow \kappa_e(N_F,N) = 2\alpha_e (N_F - 11N/2)/3\pi$.
In the limit when the fermion mass $m^2\ll |{\cal R}|$, where 
${\cal R}$ is the curvature scalar, one can partially
resum~\cite{ParkerToms:1985} terms involving the
curvature scalar in the Schwinger-DeWitt 
series~\cite{CalzettaJackParker:1985,MazzitelliSpedalieri:1995}. 
For example, the effective action of a non-Abelian gauge 
theory~\cite{CalzettaJackParker:1985} contains the following logarithmic 
term 
\begin{equation}
{\cal S}_{\rm ln} = \int d^{\,4} x (-g)^{\frac{1}{2}}\;
\left[  -\frac{1}{4}
   g^{\mu\nu}g^{\rho\sigma}F^a_{\mu\nu}F^a_{\rho\sigma}
  \left(1-\frac{11}{12}\frac{\alpha_e C}{\pi}
     \ln\left(-\frac{{\cal R}}{24\pi\mu^2}\right)\right)
  \right],
\label{magf.31b}
\end{equation}
where $\mu$ is a mass scale,  $\alpha_e=e^2/4\pi$, $F^a_{\mu\nu}$ is the 
gauge field strength, and $C$ is the defined 
in terms of the structure constants by 
$tr[T^aT^b]= - f^{acd}f^{bcd} = -C\delta^{ab}$,
such that $C=N$ for $SU(N)$. When the curvature in~(\ref{magf.31b}) 
is large, the effective gauge coupling becomes small, and one encounters
a curvature induced asymptotic freedom. Now, since in inflation
${\cal R} = -(1-3w)\rho/M_P^2$ ({\it cf.} Eq.~(\ref{magf.A4})) and 
$\rho\propto a^{-3(1+w)}$, we have 
${\cal R}^{-1} d{\cal R}/d\tau = -3(1+w) da/ad\tau$. This then implies
that one can include the effect of the logarithmic curvature running 
from the non-Abelian loop corrections in inflation by the replacement 
$\kappa_e\rightarrow \kappa_e + (1+w)(11\alpha_e C/4\pi)$ in 
Eq.~(\ref{magf.32}) ({\it cf.} Eq.~(\ref{magf.33})). Similarly, for 
scalar electrodynamics~\cite{MazzitelliSpedalieri:1995} we have 
$\kappa_e\rightarrow \kappa_e + (1+w)(\alpha_e N_s/8\pi)$,
where $N_s$  is the number of light scalar particles  
($m_s^2\ll |{\cal R}|$). With these remarks, the analysis presented in 
this section applies also to the resummed scalar
curvature corrections. Note that in de Sitter inflation $w=-1$, 
and hence the logarithmic corrections in~(\ref{magf.31b}) do not lead
to any amplification of gauge fields in inflation. 
 
We will now show that, just as in the case of the Drummond-Hathrell action,
in inflation the gauge fields are amplified in presence of massless fermions,
and damped in presence of massless bosons. Making use of the variational 
principle, $\delta {\cal S}_{\rm anom}/\delta A_\nu = 0$, we arrive
at the following equation of motion for the transverse gauge field 
\begin{equation}
\left(\partial_\tau^2 - \nabla ^{\;2} +
  \kappa_e \frac{\partial_\tau a}{a}\partial_\tau\right) 
        A_i^{\,T} = 0 \,.
\label{magf.33}
\end{equation}
We now quantize the field as in equation~(\ref{magf.6}) and get the 
following mode equation
\begin{equation}
\left(\partial_\tau^2 +\vec k^{\;2} +
  \frac{\kappa_e}{2} \left(1 - \frac{\kappa_e}{2}\right)
    \left(\frac{\partial_\tau a}{a}\right)^2 - 
      \frac{\kappa_e}{2} \frac{\partial_\tau^2 a}{a}\right) 
        {\cal A}_{\vec k} = 0 \,
\label{magf.34}
\end{equation}
with the conformal rescaling 
\begin{equation}
  \zeta = \frac{\kappa_e}{2} \equiv \frac{\alpha_e N_F}{3\pi}\,.
\label{magf.35}
\end{equation}
Assuming a power law inflation with the scale
factor $a\propto \tau^{2/(1+3w)}$ (see Eq.~(\ref{magf.16})), this becomes 
\begin{equation}
\left(\partial_\tau^2 +\vec k^{\;2} -
  \frac{\kappa_e^\prime(1+\kappa_e^\prime)}{\tau^2}\right) 
        {\cal A}_{\vec k} = 0 \,,\qquad 
  \kappa_e^\prime = -\frac{\kappa_e}{1+3w}\,,
\label{magf.36}
\end{equation}
such that $\kappa_e^\prime$ can be identified with $\theta_w$ of 
equation~(\ref{magf.9}), which is also
consistent with equation~(\ref{magf.10}). This than means that 
the mode functions in inflation are solved by equation~(\ref{magf.11})~:
\begin{equation}
{\cal A}^{(j)}_{\vec k} = \frac{1}{2} \sqrt{-\pi\tau}
   \;H^{(j)}_{\nu_i}(-k\tau)\,,\quad j=1,2,\qquad
          \nu_i =  \frac{1}{2} + \kappa_e^\prime \,,
\label{magf.37}
\end{equation}
which satisfy the Wronskian~(\ref{magf.13}). This is consistent with
the canonical quantization: \break
$\bigl[A_i(\tau,\vec x\,), \Pi_j(\tau,\vec y\,)\bigr]$
$= i\delta_{ij} \delta(\vec x - \vec y\,)$, where  
$\Pi_j = a^{\kappa_e}\partial_\tau A_j$ denotes the canonical momentum. 

\subsection{Photon in radiation era}
\label{Photon in radiation era}

An important difference between the Dolgov anomaly and the Drummond-Hathrell
action is that the anomaly also survives in radiation and matter eras,
since it is does not dependent on the energy-density scaling. 
As a consequence the matching conditions on the mode functions 
now differ from Eqs.~(\ref{magf.15}) :
\begin{eqnarray}
{\cal A}_{\vec k}^{(1)}(\tau_i) &\equiv& {\cal A}_0 =
   \alpha_{\vec k}{\cal A}_{\vec k}^{(+)}(\tau_r)
   +  \beta_{-\vec k}^*{\cal A}_{\vec k}^{(-)}(\tau_r)
\nonumber\\
\partial_\tau {\cal A}_{\vec k}^{(1)}(\tau_i)
    &\equiv& - {\cal E}_0 = 
  \alpha_{\vec k}\partial_\tau{\cal A}_{\vec k}^{(+)}(\tau_r)
   +  \beta_{-\vec k}^*\partial_\tau{\cal A}_{\vec k}^{(-)}(\tau_r) ,
\label{magf.38}
\end{eqnarray}
where the radiation era modes are the following spherical Bessel 
functions
\begin{equation}
  {\cal A}_{\vec k}^{(+)}(\tau) = 
  {\cal A}_{\vec k}^{(-)\,*}(\tau) = 
\frac{1}{2}\sqrt{\pi\tau}
   \;H^{(2)}_{\nu_r}(k\tau)\stackrel{\tau\rightarrow\infty}{\longrightarrow}
   \frac{1}{\sqrt{2k}}e^{-i(k\tau-\pi\nu_r/2-\pi/4)}
\,,\qquad \nu_r = \frac{1 - \kappa_e}{2}\,.
\label{magf.39}
\end{equation}
The following parametrization of the scale factor is convenient
({\it cf.} Eqs.~(\ref{magf.16}) -- (\ref{magf.17}))
\begin{equation}
a(\tau) =
\cases
{(-H\tau)^{{2}/{(1+3w)}}\,,\qquad & $\tau\leq \tau_i=-{1}/{H}$
 \cr
a(\tau)= \frac{\tau}{\tau_r}\,,\qquad &
  $\tau\geq \tau_r=-\frac{1+3w}{2}\frac{1}{H}$ \,.
}
\label{magf.40}
\end{equation}
We now make use of the Wronskian 
${\bf W}[{\cal A}_{\vec k}^{(+)},  {\cal A}_{\vec k}^{(-)}] = i$, and 
arrive at 
\begin{eqnarray}
  \alpha_{\vec k} &=& 
  -i \left[
  {\cal A}_{\vec k}^{(1)}(\tau_i)\partial_\tau {\cal A}_{\vec k}^{(-)}(\tau_r) 
     - (\partial_\tau{\cal A}_{\vec k}^{(1)}(\tau_i))
       {\cal A}_{\vec k}^{(-)}(\tau_r) 
\right]
\nonumber\\
  \beta_{-\vec k}^* &=& 
  i \left[
  {\cal A}_{\vec k}^{(1)}(\tau_i)\partial_\tau {\cal A}_{\vec k}^{(+)}(\tau_r) 
     - (\partial_\tau{\cal A}_{\vec k}^{(1)}(\tau_i))
       {\cal A}_{\vec k}^{(+)}(\tau_r) 
\right]
\label{magf.41}
\end{eqnarray}
which, with the help of equation~(\ref{magf.14}), reduce to 
\begin{eqnarray}
  \alpha_{\vec k} &=&  - \beta_{-\vec k} = 
  - \frac{1}{2}\left(-\frac{1+3w}{2}\right)
^{\nu_r-1/2} 
\Gamma(\nu)\left(\frac{1}{\Gamma(\nu_r)} +
  \frac{i\cos\pi\nu_r  \Gamma(1-\nu_r)}{\pi}  \right)
 \left(\frac{2H}{k}
\right)^{\nu_i-\nu_r} 
\nonumber\\
 &+&  o(k^{\nu_i-\nu_r})\,.
\label{magf.42}
\end{eqnarray}
For $\kappa_e>0$ this is the dominant term since in this case 
$\nu_i-\nu_r = -(1-3w)\kappa_e/2(1+3w) > 0 $.
Note that amplification maximizes in de Sitter inflation, when
$\alpha_{\vec k}=-\beta_{-\vec k} \propto k^{-\kappa_e}$.
After some algebra we obtain for the gauge field in radiation
\begin{eqnarray}
  {\cal A}_{\vec k}^{\rm rad}(\tau) &=& {\cal A}_0^{(\rm an)}
 \cos\left(k\tau-\frac{\pi}{2}(\nu_i-1/2)\right)
\nonumber\\
{\cal A}_0^{(\rm an)} &=& 
 -\frac{i}{\sqrt{2k}}\left(-\frac{1+3w}{2}\right)^{\frac{1}{2}-\nu_i}
 \frac{\Gamma\left(\nu_i\right)\Gamma\left(1-\nu_r\right)}{\pi}
 \left(\frac{2H}{k}
\right)^{\nu_i-\nu_r}\,,
\label{magf.43}
\end{eqnarray}
where we used equations~(\ref{magf.39}) and~(\ref{magf.42}).
For de Sitter inflation this reduces to
\begin{equation}
  {\cal A}_{\vec k}^{\rm rad}(\tau) \stackrel{deSitter}{\longrightarrow} 
 -\frac{i}{\sqrt{2k}}
 \frac{\Gamma\left(\frac{1 + \kappa_e}{2}\right)^2}{\pi}
 \left(\frac{2H}{k}
\right)^{-\kappa_e} \cos(k\tau + \pi\kappa_e/4)\,.
\label{magf.44}
\end{equation}
When conductivity grows quickly in radiation era to a large value, the smooth
matching~(\ref{magf.38}) is not any more appropriate. In this case
the field amplitude freezes-out to the value at the end of 
inflation~(\ref{magf.37}) implying that, at the beginning of radiation era,  
\begin{equation}
  {\cal A}_{\vec k}^{\rm rad}(\tau_r)
  \stackrel{\sigma\rightarrow\infty}{\longrightarrow} 
 -\frac{i}{\sqrt{2\pi k}}\Gamma\left(\nu_i\right)
 \left(\frac{2H}{k}\right)^{\nu_i-1/2}\,.
\label{magf.45}
\end{equation}
Note that this amplitude is smaller than~(\ref{magf.43}),
since $(\nu_i-\nu_r)-(\nu_i-1/2) = \kappa_e/2 >0$.

\subsection{Magnetic field spectrum from the Dolgov anomaly}
\label{Magnetic field spectrum from the Dolgov anomaly}

We first consider the simpler case~(\ref{magf.45}), for which  
the magnetic field gets frozen in the plasma at the beginning of radiation
era. The volume-averaged spectrum can be computed as in
section~\ref{Magnetic field spectrum}, resulting in the following spectrum 
today
\begin{eqnarray}
B_{\ell_c,\,\kappa_e} &=& \frac{2^{(\kappa_e-1)/2}}{\pi^{1/2}}
  \Gamma\left(\frac{\kappa_e+1}{2}\right) b_{\frac{\kappa_e + 1}{2}}
\left(\frac{T_0}{T_{\rm H}}\right)^\frac{\kappa_e}{2}
\frac{H^\frac{\kappa_e}{2}}{\ell_c^{\,2-\frac{\kappa_e}{2}}}\,
\nonumber\\
 b_{\frac{\kappa_e + 1}{2}}^2 &=& \frac{9\times 2^{\kappa_e-8}}{\pi^{3/2}}
\frac{\Gamma\left(2 -\kappa_e/2\right)\Gamma\left(\kappa_e/2\right)}
   {\Gamma\left((\kappa_e + 1)/2\right)\Gamma\left(2+\kappa_e/2\right)}
\,,
\label{magf.46}
\end{eqnarray}
where for simplicity we assumed de Sitter inflation, for which the 
resulting field is strongest.
The scaling~(\ref{magf.35}) introduces ambiguity to the field amplitude
normalization. Our normalization corresponds to setting the scale factor
$a(\tau_i)=1$ at the end of inflation, just as in equation~(\ref{magf.40}),
such that at $\tau=\tau_i$ the horizon scale modes have the amplitude
of the order the standard Minkowski vacuum amplitude. 
We believe that this justifies our normalization prescription.
The analysis of the spectrum~(\ref{magf.46}) is identical to that of 
the Drummond-Hathrell spectrum~(\ref{magf.24}) discussed in 
equations~(\ref{magf.25}) --~(\ref{magf.28}), with the replacement 
$\theta_w\rightarrow \kappa_e/2$ and $M=H$.  

The second case~(\ref{magf.43}) --~(\ref{magf.44}) applies when
the conductivity is low in `radiation era'. This means that the inflaton 
oscillates for a long time before charged particles are produced that 
freeze-out the magnetic field. Denoting the freeze-out temperature 
by $T_{\rm freeze}\ll T_H$, we then find that 
the gauge field dynamics in de Sitter inflation~(\ref{magf.44}) 
induces the following the spectrum today    
\begin{eqnarray}
B_{\ell_c,\,\kappa_e} &=& \frac{2^{-\kappa_e-1/2}}{\pi}
  \Gamma\left(\frac{\kappa_e+1}{2}\right)^2 b_{\kappa_e +\frac 12}
\left(\frac{T_0}{T_{\rm freeze}}\right)^{\kappa_e}
\left(\frac{T_{\rm freeze}}{T_H}\right)^{3\kappa_e/2}
\frac{H^{\kappa_e}}{\ell_c^{\,2-\kappa_e}}\,
\nonumber\\
 b_{\kappa_e +\frac 12}^2 &=& \frac{9\times 2^{2\kappa_e-8}}{\pi^{3/2}}
\frac{\Gamma\left(2 -\kappa_e\right)\Gamma\left(\kappa_e\right)}
   {\Gamma\left(\kappa_e + 1/2\right)\Gamma\left(\kappa_e + 2\right)}
\,.
\label{magf.47}
\end{eqnarray}
There is subtle difference between this spectrum and~(\ref{magf.24}). 
The additional suppression here 
is due to the anomalous scaling $a_{\rm freeze}^{-\kappa_e/2}
=\left({T_{\rm freeze}}/{T_H}\right)^{\kappa_e/2}$ of the gauge field.
Here $T_{\rm freeze}$ denotes the temperature at which the modes
freeze out due to a large conductivity. 
This expression is correct for the modes that enter horizon before
the freeze-out. We do not discuss here the modes that grow for a while 
and freeze out before reentering horizon in radiation era. The dynamics of 
these modes is quite complex and their spectra lie in between
those given in equations~(\ref{magf.46}) and~(\ref{magf.47}). 

The physics of the inflaton decay may in fact be 
nonperturbative, and there is a large literature on the inflaton decay and 
preheating after inflation. Nevertheless, the gauge field dynamics 
at preheating is not yet fully understood~\cite{Bassett-Rajantie}.
A heuristic account of preheating with gauge fields is given in 
Ref.~\cite{DavisDimopoulosProkopecTornkvist-II}.

 The magnetic field strength corresponding to the spectra~(\ref{magf.46})
and~(\ref{magf.47}) is very similar to those plotted in figure~\ref{spectra}
and hence we do not plot them here. An interesting question is of course 
under what conditions~(\ref{magf.46}) and~(\ref{magf.47}) can result in 
spectra with a slope $\gsim -1$, required to seed the galactic dynamo. 
To get the spectrum $B_\ell\propto \ell^{-1}$, equations~(\ref{magf.46})
and~(\ref{magf.47}) imply $1\leq \kappa_e \leq 2$, or equivalently 
\begin{equation}
\frac{3\pi}{2\alpha_e} \leq   N_F \leq \frac{3\pi}{\alpha_e}\,.
\label{magf.48}
\end{equation}
For $\alpha_e\sim 1/40$ this gives the following bound
\begin{equation}
2\times 10^2 \lsim   N_F \lsim 4\times 10^2
\label{magf.49}
\end{equation}
for the number of charged fermions. In presence of bosonic fields
amplification is reduced. For example, when 
the electromagnetism is embedded in an SU(N), then 
equation~(\ref{magf.48}) should be read as a limit on $N_F-11N/2$. 
When applied to the standard model, the photon field is amplified 
on scales smaller than the electroweak scale $M_W\sim 80$~GeV. 
On higher scales, the hypercharge field is amplified, and the relevant  
number of light fermions is $N_F=12$ (above the top mass). 
Hence this mechanism results in weak magnetic field
amplification if the standard model fermions are the only light fermions. 
To get strong amplification, one would need a few hundred of additional
fermions with masses at some intermediate scale below inflation. 
It is not clear whether this can be made consistent with  
the grand unification.

 The validity of the Dolgov anomaly equation~(\ref{magf.33}) was 
questioned in~\cite{MazzitelliSpedalieri:1995}. Here we have solved
Eq.~(\ref{magf.33}) exactly to order $e^2$. To properly
address the question of validity of this approach one would have
to study the higher order corrections in the coupling constant 
expansion to the trace anomaly.

\section{Photon coupling to scalar and pseudoscalar fields}
\label{Photon coupling to scalar and pseudoscalar fields}

\subsection{Photon coupling to scalar field}
\label{Photon coupling to scalar field}

Consider now the following action with gauge invariant coupling 
of the gauge field to a scalar field
\begin{equation}
{\cal S}_{\phi} 
 = \int d^4 x \sqrt{-g}\;\left(
 - \frac{1}{4} f(\phi) g^{\mu\rho}g^{\nu\sigma}F_{\mu\nu}F_{\rho\sigma}
 + \frac 12 g^{\mu\nu}(\partial_\mu \phi)(\partial_\nu \phi) 
 +  V(\phi)
\right),
\label{magf.50}
\end{equation}
where $F_{\mu\nu}$ is the gauge field strength, $\phi$ is a scalar field
$\phi$ and $V(\phi)$ is the corresponding potential. 
In general $\phi$ may couple to other fields as well. The `coupling'
function $f=f(\phi)$ we take to be of the form: 
\begin{equation}
 f = \sum_{n=0}^\infty f_n \left(\frac{\phi}{M_P}\right)^n 
,
\label{magf.51}
\end{equation}
where $f_n$ are some coupling coefficients
and $M_P = (8\pi G)^{-1/2}\approx 2.4\times 10^{18}$GeV is the reduced 
Planck mass. For the standard gauge kinetic term, $f_0=1$, and in
addition the potential $V(\phi)$ must be chosen such that today $\phi=0$
and  $V(\phi=0)=0$. A particular realization of this model has
been considered by Ratra~\cite{Ratra:1992bn}, where $f(\phi)\propto
e^{-\lambda\phi/M_P}$, and $\phi$ was taken to be the inflaton in an
extended inflation with the potential
$V(\phi)\propto e^{-\lambda'\phi/M_P}$. 
In this case $f_n=(-\lambda)^n/n!$. Ratra's model has subsequently been
reconsidered in the context of string-inspired
cosmology~\cite{Gasperini:1995dh,Lemoine:1995vj}, where the role of
$\phi$ is taken by the dilaton field. Here we study the conditions
under which $f(\phi)$ and $V(\phi)$ can lead to significant 
amplification of the magnetic field in inflation. As a particular case 
we discuss Ratra's model. 

Working in Coulomb gauge, $A_0=0$ and $\nabla \cdot \vec A =0$
({\it cf.} Eqs.~(\ref{magf.2}-\ref{magf.3})), the action~(\ref{magf.50})
implies the following equation of motion in de Sitter inflation 
for the transverse gauge field
\begin{equation}
  \left(\partial_\tau^2  - \nabla^2
      + 
\frac{d\phi}{d\tau} \left (\partial_\phi \ln f(\phi) \right)\partial_\tau
  \right) A_i^{\,T}  
      = 0 
,
\label{magf.52}
\end{equation}
where we assumed that in inflation the field $\phi=\phi(\tau)$ is 
a function of time only. It is now clear that the coupling function $f$ 
is responsible for breakdown of conformal invariance, and hence its
precise form determines whether the field gets amplified in inflation.
In what follows we consider two types of functions: (A) quadratic 
coupling function
\begin{equation}
 f =  1 + f_2 \frac {\phi^2}{M_P^2} +.. \qquad\quad {\rm (Type\; A)}
,
\label{magf.53}
\end{equation}
and (B) exponential coupling function
\begin{equation}
 f =  e^{-\lambda\phi/M_P}\qquad\quad {\rm (Type\; B)}
.
\label{magf.54}
\end{equation}
{\it Type A} coupling functions are expected to be induced by the 
Planck physics. The coupling $f_1$ can be made to vanish
in~(\ref{magf.53}) by imposing the symmetry $\phi\rightarrow -\phi$. 
Similarly, by promoting $\phi$ to a complex field and imposing
a $Z_N$ symmetry, one gets that $f_n=0$, $\forall n< N$.
{\it Type B} coupling functions may occur from string theory corrections
to the gauge field dynamics.

\subsubsection{Chaotic inflation}

Making use of equation~(\ref{magf.c6}) in Appendix C.1 where we 
estimate $\partial_\tau \phi$ during slow-roll in chaotic inflation,
where $V = m^2 \phi^2/2$, we can recast equation~(\ref{magf.52}) as
\begin{equation}
  \left(\partial_\tau^2  - \nabla^2
      + \frac{2M_P^2}{\phi}\frac {1}{\tau}
        \left (\partial_\phi \ln f(\phi) \right)\partial_\tau
  \right) A_i^{\,T}  
      = 0.
\label{magf.55}
\end{equation}
For {\it Type A} coupling functions this implies
\begin{equation}
  \left(\partial_\tau^2  - \nabla^2
      - \frac {2\theta_2}{\tau} \partial_\tau
  \right) A_i^{\,T}
      = 0    ,\qquad \theta_2 =  - 2f_2
,
\label{magf.56}
\end{equation}
where we neglected the higher order corrections. Similarly, for
{\it Type B} coupling functions we obtain the following equation:
\begin{equation}
  \left(\partial_\tau^2  - \nabla^2
      - \frac {2\theta_\lambda}{\tau} \partial_\tau
  \right) A_i^{\,T}
      = 0    ,\qquad \theta_\lambda =  \lambda\frac{M_P}{\phi}
,
\label{magf.57}
\end{equation}
where $\phi = \phi_0 + (2M_P^2/\phi) \ln(-H_0\tau)$. Since
$\theta_2$ and $\theta_\lambda$ are generally slowly varying functions
of conformal time, equations~(\ref{magf.56}-\ref{magf.57}) can be well
approximated by the Bessel equation~(\ref{magf.7}) analysed in detail in 
section~\ref{Drummond-Hathrell action}. The corresponding magnetic field 
spectrum is calculated in~(\ref{magf.23}-\ref{magf.24}), according to which 
the spectrum reads
\begin{equation}
B_{\ell_c}\propto \ell_c^{-2+\theta}, \qquad (\theta=\theta_2,\theta_\lambda)
,
\label{magf.58}
\end{equation}
where $\ell_c$ is the comoving scale. To get the amplitude 
one should set $M = H$ in~(\ref{magf.23}-\ref{magf.24}), which 
corresponds to the normalization {\it to} the Minkowski
vacuum at the horizon crossing. We assume that this normalization
is effected by the local interactions in the de Sitter vacuum operative
on subhorizon scales. With this the magnetic field spectrum can be 
simply read off from figure~3. To get an acceptable spectrum, we then have 
$2 \geq \,  \theta_{2}, \theta_{\lambda} \,\geq 1$, or 
\begin{equation}
  \frac 12 \;\leq\;   -f_{2} \;\leq\; 1,
\qquad\quad
  \lambda \sim 30 
,
\label{magf.59}
\end{equation}
where we took $M_P/\phi\simeq 1/(2\times 6^{1/4} N^{1/2})\sim 1/20$, and 
the number of $e$-folds $N\sim 50$. We have thus found that one can obtain
magnetic field spectra with observable consequences
with $f_2$ of order unity, which is a natural value.
A generalization of {\it Type A} model contains 
higher order terms in the $\phi/M_P$-expansion. 
Since in chaotic inflation typically $\phi\gg M_P$, the resulting
magnetic field spectra of interest can be obtained with coupling
coefficients $|f_{2n}|\ll 1$ ($n=2,3, ..$), where $f_0=1$, $f_1=f_2=f_3=0$. 
On the other hand, for {\it Type B} model the coupling $\lambda$ is 
required to be unnaturally strong. Moreover, one gets a significant 
amplification in inflation only when the effective gauge coupling 
$e^2(\phi) = e^2_0\, e^{-\lambda\phi/M_P}$ starts very weak in inflation. 

In the spectrum computation in section~\ref{Magnetic field spectrum}
we have neglected the conductivity in radiation era. A simple way of 
accounting for that effect is to assume that magnetic field freezes in when
the conductivity becomes large. The spectrum 
calculation in this case can be performed analogously as in 
section~\ref{Magnetic field spectrum from the Dolgov anomaly} 
({\it cf.} Eq.~(\ref{magf.46})). To get spectra of physical interest 
in this case, one requires by about a factor of two larger values of
$\theta_{2}$ and $\theta_{\lambda}$ from those indicated in~(\ref{magf.59}). 

\subsubsection{Extended inflation}

For the potential of extended inflation $V = V_0 e^{-\lambda'\phi/M_P}$
equation~(\ref{magf.52}) can be written as
\begin{equation}
  \left(\partial_\tau^2  - \nabla^2
      - \frac{2M_P}{\lambda'\left(\frac{2}{{\lambda'}^{2}} -1\right) }
          \frac {1}{\tau}
        \left (\partial_\phi \ln f(\phi) \right)\partial_\tau
  \right) A_i^{\,T}  
      = 0.
\label{magf.60}
\end{equation}
where, to estimate $d\phi/d\tau$, we made use of 
equation~(\ref{magf.c14}) in Appendix C.2. 
For {\it Type A} model~(\ref{magf.53}) this implies
\begin{equation}
  \left(\partial_\tau^2  - \nabla^2
      - \frac {2\theta_2'}{\tau} \partial_\tau
  \right) A_i^{\,T}
      = 0    ,\qquad \theta_2' =  
 \frac{2f_2}{\lambda'\left(\frac{2}{{\lambda'}^{2}} -1\right)}\frac{\phi}{M_P}
,
\label{magf.61}
\end{equation}
where $\phi$ is a slowly varying function of conformal time given 
in~(\ref{magf.c15}). Since
$\phi\sim\phi_0\sim -(M_P/\lambda')\ln (12/{\lambda'}^4)$, 
this immediately implies 
\begin{equation}
1\lsim \theta_2'\lsim 2 \quad\Longrightarrow\quad   
  f_2 \sim - \left(\ln\frac{12}{{\lambda'}^4}\right)^{-1}
   \qquad (\lambda'\ll 1),
\label{magf.62}
\end{equation}
which is to be compared with the conditions in chaotic 
inflation~(\ref{magf.57}) and~(\ref{magf.59}). 
For {\it Type B} model~(\ref{magf.54}) 
we get the following equation:
\begin{equation}
  \left(\partial_\tau^2  - \nabla^2
      - \frac {2\theta_\lambda'}{\tau} \partial_\tau
  \right) A_i^{\,T}
      = 0    ,\qquad \theta_\lambda' =  -
  \frac{\lambda}{\lambda'\left(\frac{2}{{\lambda'}^{2}} -1\right)}
\,.
\label{magf.63}
\end{equation}
To get a magnetic field of observable strength we than 
have
\begin{equation}
1\lsim \theta_\lambda' \lsim 2  \quad\Longrightarrow\quad 
\frac{2}{\lambda'} \lsim -\lambda \lsim \frac{4}{\lambda'} 
\qquad (0<\lambda'\ll 1)
\,.
\label{magf.64}
\end{equation}
In this case the effective gauge coupling 
$e^2(\phi)=e^2_0 e^{-\lambda\phi/M_P}$ begins very weak at early stages 
in inflation ($\phi\ll -M_P$) and acquires today's value at the end
of inflation ($\phi \approx 0$). Precisely this feature of the 
pre-Big Bang model was used in~\cite{Gasperini:1995dh} to get
magnetic field amplification, where the gauge coupling constant 
is determined dynamically by the dilaton expectation value.

 An interesting question is whether there is a natural spectrum 
predicted by inflation. In other words, is there a value of the spectrum
slope $2-\theta$ that is in any way singled out.
Assume that at early stages of inflation $\theta>2$, which 
implies a (tachyonic) instability resulting in a fast amplitude
growth and spontaneous magnetization on superhorizon scales. 
This condensate may affect evolution of the fields in inflation
and possibly single out the scale invariant spectrum. 
This question deserves further investigation.

\subsection{Photon coupling to pseudoscalar field}
\label{Photon coupling to pseudoscalar field}

 We now consider the following coupling of a pseudoscalar field 
the Chern-Pontryagin density
\begin{equation}
{\cal S}_{\psi} =\int d^4 x \sqrt{-g}\;\left(
  -\frac{1}{4} g^{\mu\rho}g^{\nu\sigma}F_{\mu\nu} F_{\rho\sigma}
  -\frac{1}{4} h(\psi) g^{\mu\rho}g^{\nu\sigma}F_{\mu\nu}\tilde F_{\rho\sigma}
  + \frac 12 g^{\mu\nu}(\partial_\mu \psi)(\partial_\nu \psi) 
  + V(\psi)
\right)
\label{magf.65}
\end{equation}
where 
\begin{equation}
    h = \sum_{n=0}^\infty h_n \left(\frac{\phi}{M_P}\right)^n 
\label{magf.66}
\end{equation}
is a coupling function of a pseudoscalar field $\psi$, 
$\tilde F_{\mu\nu} = \epsilon_{\mu\nu\eta\rho}F_{\eta\rho}/2$ is the
dual field strength and $\epsilon_{\mu\nu\eta\rho}$ is the totally
antisymmetric tensor defined by $\epsilon_{0123}=1$. 
The case $h(\psi)=h_2\psi^2/M_P^2$ is mentioned in~\cite{TurnerWidrow}
as potentially interesting, but no attempt was made to solve it. 
There is an observational constraint on $h$ coming from the magnetic 
dipole moment measurements of the neutron and electron, according to
which $|h_0|\lsim 10^{-9}$ ($\psi=0$); for simplicity here we assume
$h_0=0$ and $\psi=0$. With $\psi=0$ today there are no experimental 
constraints on $h_n$ ($n\geq 1$). When applied to the quantum 
chromodynamics, the question of smallness of $h$ is termed the strong
CP-problem. Peccei and Quinn have suggested that $h$ might be
dynamically driven to zero by symmetry breaking of $\psi$ at some
high energy scale. The special case of this 
model was considered in~\cite{GarretsonFieldCarroll:1992},
where the authors assumed that the axionic potential is of the form
$V(\psi)=V_0[1-\cos(\psi/\psi_0)]$, and found no significant amplification. 

The action~(\ref{magf.65}) implies the following equation of motion
for the gauge field in inflation in Coulomb gauge
({\it cf.} Eq.~(\ref{magf.52})):
\begin{equation}
  \left(\partial_\tau^2  - \nabla^2
      + 
\frac{d\psi}{d\tau} \left (\partial_\psi \ln h(\psi) \right)\nabla\times
  \right) A_i^{\,T}  
      = 0 
,
\label{magf.67}
\end{equation}
where we used $\eta^{\mu\rho}\partial_\mu \tilde F_{\rho\nu}=0$ and 
assumed that in inflation $\psi=\psi(\tau)$. Conformal invariance 
is now broken by $h=h(\psi)$, which then may lead to magnetic field 
amplification in inflation. It now follows from the analysis in 
section~\ref{Photon coupling to scalar field}, that in inflation 
quite generically equation~(\ref{magf.67}) can be rewritten as 
\begin{equation}
  \left(\partial_\tau^2  - \nabla^2
      - \frac{2\theta_\psi}{\tau}\nabla\times
  \right) A_i^{\,T}  
      = 0 
,
\label{magf.68}
\end{equation}
where $\theta_\psi$ is a slowly varying function of time in inflation,
and $\theta_\psi\approx 0$ in radiation era. The specific forms 
of $\theta_\psi$ for {\it Type A} and {\it B} 
models~(\ref{magf.53}-\ref{magf.54}) in chaotic and extended inflation
can be easily reconstructed from
Eqs.~(\ref{magf.56}-\ref{magf.57}), (\ref{magf.61}) and~(\ref{magf.63}),
provided $f\rightarrow h$ and  $\phi\rightarrow \psi$.

Equation~(\ref{magf.68}) can be written in terms of the mode 
functions~(\ref{magf.6}) as follows
\begin{equation}
  \left(\partial_\tau^2  + k^2\right) \epsilon^T_{\vec k} {\cal A}^T_{\vec k}
-\frac{2\theta_\psi}{\tau}i\vec k\times \epsilon^T_{\vec k}{\cal A}^T_{\vec k}
      = 0 
.
\label{magf.69}
\end{equation}
Since the circular polarization vectors satisfy 
$i\hat k\times \epsilon^\pm_{\vec k} = \pm \epsilon^\pm_{\vec k}$
($T=\pm$)
we can write~(\ref{magf.69}) for the {\it circularly polarized}
mode amplitudes as
\begin{equation}
  \left(\partial_\tau^2  + k^2 \mp\frac{2\theta_\psi}{\tau} k\right) 
    {\cal A}^\pm_{\vec k}  = 0 
.
\label{magf.70}
\end{equation}
Note that when $\theta_\psi<0$ ($\theta_\psi>0$) 
the mode ${\cal A}_{\vec k}^{+}$ 
(${\cal A}_{\vec k}^{-}$) gets amplified in inflation (where we took into
account that $\tau<0$ in inflation), such that in principle one can
distinguish the magnetic fields produced by a pseudoscalar coupling
to the Chern-Pontryagin density from other mechanisms discussed
in this paper, where the field amplification is polarization independent.
This is of course true provided amplification is strong
enough to be observable. When written in terms of the variable $u=2ik\tau$,
equation~(\ref{magf.70}) reduces to the following Whittaker equation 
\cite{GradshteynRyzhik}:
\begin{equation}
  \left(\partial_u^2  - \frac 14  \pm \frac{i\theta_\psi}{u}\right)
   {\cal A}_{\vec k}^{\pm}
      = 0 
.
\label{magf.71}
\end{equation}
%
The solutions can be expressed in terms of
the Whittaker functions $W_{\frac 12,\lambda}(u)$ as
%
\begin{equation}
   {\cal A}_{\vec k}^{\pm}(\tau)
  = \frac{e^{\mp\frac\pi 2\theta_\psi}}{\sqrt{2k}} 
  \; W_{\frac 12,\, \pm i\theta_\psi}(2ik\tau) 
,
\label{magf.72}
\end{equation}
with the Wronskian normalization 
${\bf W}[{\cal A}_{\vec k}^{\pm}(\tau),{\cal A}_{\vec k}^{\mp}(-\tau)]=i$,
where ${\cal A}_{\vec k}^{\mp}(-\tau)$ denotes the (second) linearly 
independent solution. The asymptotic form of~(\ref{magf.72}) is 
\begin{equation}
   {\cal A}_{\vec k}^{\pm}(\tau)
   \quad\stackrel{k|\tau|\rightarrow\infty}{\longrightarrow}\quad
      \frac{1}{\sqrt{2k}} 
       e^{-ik\tau \pm i\theta_\psi\ln(-2k\tau)}
,
\label{magf.73}
\end{equation}
such that it approaches the Minkowski vacuum at $\tau\rightarrow -\infty$. 
The only effect of the $\psi$-field at a distant past is a time-dependent
phase shift in the mode functions. For superhorizon modes close to the end
of inflation the following asymptotic form is useful
\begin{equation}
   {\cal A}_{\vec k}^{\pm}
   \quad\stackrel{k|\tau|\rightarrow 0}{\longrightarrow}\quad
      \frac{1}{\sqrt{2k}} \,
       \frac{e^{\mp\frac\pi 2\theta_\psi}}{\Gamma(1\pm i\theta_\psi)}
.
\label{magf.74}
\end{equation}
Now matching this smoothly onto the radiation vacuum solutions
${\cal A}^{\rm \pm rad}_{\vec k} = (2k)^{-1/2} e^{\mp ik\tau}$ 
results in the vacuum spectrum of a modified amplitude.
The circularly polarized modes ${\cal A}^\pm_{\vec k}$ get amplified
in inflation by a factor 
\begin{equation}
 \left|\frac{{\cal A}_{\vec k}^{\pm}}{{\cal A}^{\rm \pm rad}_{\vec k}}\right|
  \simeq e^{\frac\pi 2|\theta_\psi|\mp\frac\pi 2\theta_\psi}
,
\label{magf.75}
\end{equation}
where we made use of the Stirling formula, 
$|\Gamma(1\pm i\theta_\psi)|\simeq (2\pi|\theta_\psi|)^{1/2}
  e^{-\pi|\theta_\psi|/2}$ ($|\theta_\psi|\geq 1$). This means that,
when $\theta_\psi<0$ ( $\theta_\psi>0$), 
the mode ${\cal A}_{\vec k}^{+}$  (${\cal A}_{\vec k}^{-}$)
gets amplified by a factor $\sim e^{\pi|\theta_\psi|}$, while  
the amplitude of ${\cal A}_{\vec k}^{-}$  (${\cal A}_{\vec k}^{+}$) 
remains unchanged. 
This is an extreme case of {\it birefringence}, where the photon of
one polarization corresponds to a tachyonic particle, and the other 
to a massive vector particle. 

We have thus shown that a pseudoscalar field coupling to the
Chern-Pontryagin density~(\ref{magf.65}) cannot change the vacuum
spectrum $B_\ell \propto \ell^{-2}$ of gauge
fields in inflation. In this case conformal invariance 
is broken in inflation such that the vacuum amplitude changes 
and an unobservable phase shift to the gauge mode functions is induced.
The resulting  amplification of the vacuum spectrum is typically too weak
to be observable today. Moreover, thermal processes in radiation era 
are expected to create magnetic field with the spectrum 
$B_{\ell}^{\rm th}\propto \ell^{-3/2}$ 
~\cite{DavisDimopoulosProkopecTornkvist-II}, which dominates 
over the vacuum spectrum on scales of interest in cosmology.

 Even though this model is not relevant for cosmological magnetic fields, 
it may be of interest for baryogenesis. An explicit CP-violation in 
the pseudoscalar sector may result in a nonperturbative amplification 
of the CP-violating phase of a complex pseudoscalar field at preheating.
When the field decays into the standard model fermions, a net baryon
number may be produced. This is reminiscent of the Affleck-Dine
mechanism~\cite{AffleckDine}, with an important difference however:
the $\psi$-field is a pseudoscalar field which is not charged under the
baryon number, so that baryon violating processes are needed 
to get a nonzero baryon production. In a related 
work the dynamics of a pseudoscalar field in radiation era and 
its relevance for baryogenesis has been 
considered in~\cite{Giovannini:1998gp,Brustein:1999rk}.

\section{Photon coupling to scalar cosmological perturbations}
\label{Photon coupling to scalar cosmological perturbations}

 An interesting proposal has been recently put forward by 
Maroto~\cite{Maroto:2000zu}, who considered the effect of the 
photon field coupling to the scalar metric perturbations.
We now reconsider this mechanism
(without making a perturbative expansion of the mode amplitudes). 
In presence of the scalar cosmological perturbations the metric gets
modified as follows 
\begin{equation}
g^{\mu\nu}= a^{-2} \left((1-2\Phi)d\tau^2 - (1+2\Phi)d\vec x^{\,2}\right)
,
\label{magf.76}
\end{equation}
where $\Phi=\Phi(\eta,\vec x)$ denotes the (gauge invariant) scalar 
gravitational potential. The potential $\Phi$ is to good approximation 
scale invariant. The CMBR measurements and the large scale
structure formation imply that $\Phi\sim 10^{-5}$.

The gauge field equation of motion is simply 
\begin{equation}
\partial_\mu \left(\sqrt{-g}g^{\mu\rho}g^{\nu\sigma}F_{\rho\sigma}\right) = 0
,
\label{magf.77}
\end{equation}
which, when written in components and working 
to linear order in $\Phi$ ($|\Phi|\ll 1$), reads
\begin{equation}
\partial_i \left[(1-2\Phi)(\partial_iA_0-\partial_\tau A_i)\right] = 0
\label{magf.78}
\end{equation}
and 
\begin{equation}
\partial_\tau\left[(1-2\Phi)(\partial_iA_0-\partial_\tau A_i)\right]
+\partial_j\left[(1+2\Phi)(\partial_jA_i-\partial_i A_j)\right]
  = 0
.
\label{magf.79}
\end{equation}
Equation~(\ref{magf.78}) can be also written as 
\begin{equation}
\nabla^2 A_0 -2\nabla\Phi\cdot \nabla A_0
= \partial_\tau\nabla\cdot \vec A
  - 2(\nabla\Phi)\cdot \partial_\tau \vec A
,
\label{magf.80}
\end{equation}
where we used $\nabla \ln (1-2\Phi)\approx -2\nabla\Phi$.
Setting $A_0=0$ leads to the following generalized Coulomb gauge
\begin{equation}
\partial_\tau\nabla\cdot \vec A = 2(\nabla \Phi)\cdot \partial_\tau \vec A
.
\label{magf.81}
\end{equation}
It turns out that this modification becomes unimportant 
on superhorizon scales, where $k|\tau|\ll 1$. We can now write
equation~(\ref{magf.79}) in the gauge~(\ref{magf.81}) as
\begin{equation}
\left(\partial_\tau^2 -\nabla^2\right)A_i
    - 2(\partial_\tau\Phi)\partial_\tau A_i
    + \partial_i \nabla \cdot A
    - 2(\partial_j\Phi)(\partial_jA_i-\partial_i A_j)
    = 0
,
\label{magf.82}
\end{equation}
where we neglected $\Phi$ in favour of {\it one} 
in $1-2\Phi$ and $1+2\Phi$, which is a controlled truncation when working
to leading order in $\Phi$. The last two terms in~(\ref{magf.82}) 
can be consistently neglected, since formally they are a small
correction to the $\nabla^2A_i$-term. This follows from 
equation~(\ref{magf.81}) and $\Phi\ll 1$.
Moreover, it is easy to see that the term
$2(\partial_\tau\Phi)\partial_\tau A_i$
(which potentially leads to gauge field amplification)
dominates on superhorizon scales ($k|\tau|\ll 1$) 
over the terms we have just argued can be neglected.
With this equation~(\ref{magf.82}) simplifies to
\begin{equation}
\left(\partial_\tau^2 -\nabla^2\right)A_i
    - 2(\partial_\tau\Phi)\partial_\tau A_i
    = 0
.
\label{magf.83}
\end{equation}

 We shall now solve this equation in the {\it mean field} approximation
\begin{equation}
\Phi \rightarrow \Phi_L\equiv\langle \Phi\rangle_L
,
\label{magf.84}
\end{equation}
where $\langle \Phi\rangle_L$ denotes a spatial average over some
scale $L$. Since $\Phi$ is almost scale invariant in inflation, we 
expect $\Phi_L$ to be a slow (logarithmic) function of time and
$L$. Taking for example $\Phi_L\sim \Phi_0 \ln (-\tau/L)$
($\Phi_0\sim 10^{-5}\ll 1$), we find
$\partial_\tau\Phi\rightarrow \Phi_0/\tau$. With this we can write 
equation~(\ref{magf.83}) in the mean field approximation as
the following Bessel equation
\begin{equation}
\left(\partial_\tau^2 - \nabla^2 
    - \frac{2\Phi_0}{\tau}\partial_\tau \right)A_i
    = 0
.
\label{magf.85}
\end{equation}
In section~\ref{Drummond-Hathrell action} we study in detail this
equation ({\it cf.} Eq.~(\ref{magf.7})) and show that the solutions
for the mode functions can be written in terms of the Hankel
functions~(\ref{magf.11}) 
with the index $\nu=\frac 12 + \Phi_0$, and that the resulting mode
amplification ${\cal A}_{\vec k} \propto (-k\tau)^{\nu-1/2}$ is 
small when $|\nu-1/2|=|\Phi_0|\ll 1$. Since in radiation era
$\partial_\tau\Phi=0$, the mode functions are simply 
${\cal A}_{\vec k}^{\pm\rm rad} = ({2k})^{-1/2} e^{\mp ik\tau}$,
so that the calculation in section~\ref{Magnetic field spectrum}
leading to the spectrum~(\ref{magf.23}) applies, 
and hence we get an almost vacuum spectrum today
$B_\ell \propto \ell ^{-2+\Phi_0}$ ($\Phi_0\sim 10^{-5}$). 

We shall now argue that the mean field approximation represents 
an upper bound on the amplification on superhorizon scales.
This approximation captures the effect of a spatially averaged
gravitational potential, but does not account for inelastic 
scatterings. Since inelastic scatterings are in general dissipative,
their inclusion can lead only to suppression of gauge fields.
Moreover, the sign of $\Phi_L$ in~(\ref{magf.84}) can be both 
positive and negative, suggesting an additional suppression when
compared to~(\ref{magf.85}).   
To show that this is indeed so, we now study Eq.~(\ref{magf.83})
by making an {\it Ansatz} for the momentum dependence of the mode functions. 
Upon performing a Fourier transform on~(\ref{magf.83}) we get
\begin{equation}
\left(\partial_\tau^2 + k^2 \right){\cal A}_{\vec k}
    - 2\int \frac{d^3 k'}{(2\pi)^3}(\partial_\tau\Phi_{\vec k'})
\partial_\tau {\cal A}_{\vec k-\vec k'}
    = 0
,
\label{magf.86}
\end{equation}
where we suppressed the polarization indices. The mean field approximation 
consists of neglecting inelastic scatterings that lead to
nontrivial momentum exchange, {\it i.e.} it amounts to 
${\cal A}_{\vec k-\vec k'}\rightarrow  {\cal A}_{\vec k}$. 


 The spectrum of the potential $\Phi$ is typically close to scale invariant, 
which means that 
\begin{equation}
\Phi_{\vec k} = \phi_0(\tau)
          \Lambda^{\beta-3} k^{-\beta} 
   , \qquad k < \Lambda 
,
\label{magf.87}
\end{equation}
where $\beta$ is the spectral index for superhorizon modes ($k < \Lambda$),
and 
\begin{equation}
\phi_0 = C_0\;
 \frac{1}{a}\frac{d}{d\tau}\left(\frac{1}{a}\int a^2 d\tau\right) 
\label{magf.88}
\end{equation}
is a slowly varying function of $\tau$, and $C_0\sim 10^{-5}$ is a
normalization constant. 
For a scale invariant spectrum $\beta \equiv 5/2 - n =3/2$.
On large cosmological scales the CMBR measurements constrain
$|\beta - 3/2|\lsim 0.2$; on smaller scales a similar
constraint results from large scale structure formation. 
The cutoff $\Lambda$ in~(\ref{magf.87}) is of the order the horizon
$H$ at the end of inflation, such that 
on scales $k_{\rm phys}=k/a >\Lambda \sim H$ the vacuum spectrum
$|\Phi_{\vec k}|\sim|\phi_0|/\sqrt{k}$ is recovered. 
Similarly for the gauge field we take 
\begin{equation}
{\cal A}_{\vec k} = {\cal A}_0(\tau) k^{-\alpha} 
   , \qquad k < \Lambda 
,
\label{magf.89}
\end{equation}
where the scale invariant spectrum for the magnetic field corresponds to 
$\alpha=5/2$, and ${\cal A}_{\vec k}\propto k^{-1/2}$ for $k\geq a\Lambda$. 
Now inserting the {\it Ans\"atze} 
(\ref{magf.87}-\ref{magf.89}) into~(\ref{magf.86}) results in 
\begin{equation}
\left(\partial_\tau^2 + k^2 
     - \frac{\partial_\tau\phi_0}
       {2\pi^2(2-\beta)} \left(\frac{k}{\Lambda}\right)^{3-\beta}
       {\cal I}_{\alpha\beta}(x_0,x_1)\,\partial_\tau
       \right){\cal A}_0
     = 0,
\label{magf.90}
\end{equation}
where
\footnote{For the special case $\beta=2$ the integral 
${\cal I}_{\alpha\beta}(x_0,x_1)/(2-\beta)\rightarrow 
\int_{x_0}^{x_1} dx x^{1-\alpha}\ln \left[(1+x)/|1-x|\right]$.} 
\begin{equation}
 {\cal I}_{\alpha\beta}(x_0,x_1) = \int_{x_0}^{x_1} dx x^{1-\alpha}
\left[(1+x)^{2-\beta}-|1-x|^{2-\beta}\right]
,
\label{magf.91}
\end{equation}
where $x_1=a\Lambda/k$, and we have introduced a lower limit of integration
$x_0=k_0/k$, where $k_0$ denotes the lowest moment amplified at the
beginning of inflation. 
\begin{figure}[htbp]
\begin{center}
\epsfig{file=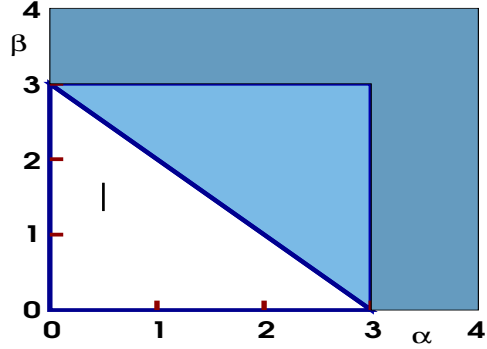, height=1.8in,width=2.5in}
\end{center}
\vskip -0.1in
\lbfig{convergence}
\caption[fig4]{%
\small The {\it unshaded} region corresponds to the phase space where 
${\cal I}_{\alpha\beta}(0,x_1=\Lambda/k)$ is dominated by 
the ultraviolet cutoff, and where the only self-consistent solutions 
of~(\ref{magf.90}) lie (marked by the {\it thin vertical line} 
$\alpha\approx 1/2$, $|\beta-3/2|\leq 0.2$).
In the {\it light blue} shaded triangle the integral
${\cal I}_{\alpha\beta}(0,\infty)$ 
in~(\ref{magf.91}) is convergent. 
}
\end{figure}
The self-consistent solution of~(\ref{magf.90}) is obtained when 
when $\alpha+\beta<3$. In this case the ultraviolet cutoff $x_1$
dominates the integral~(\ref{magf.91}), and equation~(\ref{magf.90})
becomes 
\begin{equation}
\left(\partial_\tau^2 + k^2 
     - \frac{d\phi_0/d\tau}
       {\pi^2(3-\alpha-\beta)} \left(\frac{k}{\Lambda}\right)^{\alpha}
        a^{\alpha+\beta-3}\partial_\tau
       \right){\cal A}_{\vec k}
     = 0.
\label{magf.92}
\end{equation}
Since in inflation $d\phi_0/d\tau\sim C_0/\tau$, and $a\simeq -1/H\tau$,
Eq.~(\ref{magf.92}) can be simplified to 
\begin{equation}
\left(\partial_\tau^2 + k^2 +2\gamma_\tau(k) \partial_\tau
       \right){\cal A}_{\vec k}
     = 0,\qquad
 \gamma_\tau(k) \sim \frac{C_0}{(-H\tau)^{\alpha+\beta-2}}
    \frac{1}{2\pi^2(3-\alpha-\beta)} \left(\frac{k}{\Lambda}\right)^{\alpha}
.
\label{magf.93}
\end{equation}
When $\alpha+\beta\approx 2$, $\gamma_\tau\sim \gamma_0$ is almost 
time independent, and equation~(\ref{magf.93}) is self-consistently
solved by 
\begin{equation}
{\cal A}^\pm_{\vec k} \sim \cases{ \frac{1}{\sqrt{2k}}
 e^{-\gamma_0\tau \pm i \omega_0\tau}  & when  $\gamma_0<k$  \cr
 & \cr
 \frac{1}{\sqrt{2k}} e^{-\gamma_\pm\tau} & when  $\gamma_0>k$ }
\label{magf.94}
\end{equation}
where $\omega_0= \sqrt{k^2-\gamma_0^2}$ and 
$\gamma_\pm = \gamma_0\pm \sqrt{\gamma_0^2-k^2}>0$. 
The critical case $k_c=\gamma_0(k_c)$ corresponds to 
$k_c\sim (C_0^2/4\pi^4)(H^2/\Lambda) \ll H$. In all cases 
in~(\ref{magf.94}) the spectrum on large superhorizon scales is to good 
approximation the Minkowski vacuum, and the gauge field actually decays
slowly in inflation. This  suggests that the mean field
equation~(\ref{magf.85}) can only overestimate amplification,
and in addition it indicates that the vacuum $\alpha = 1/2$ approximates 
well the actual solution. The solution to~(\ref{magf.93}) is represented 
by the thin vertical line in figure~\ref{convergence}, and it is the only
self-consistent solution of~(\ref{magf.90}). Indeed, 
when $\alpha>3$ the integral~(\ref{magf.91}) is dominated by the
infrared cutoff. In this case  
${\cal I}_{\alpha\beta}(x_0,\infty)/(2-\beta)\simeq 
(2/(\alpha-3))(k/k_0)^{\alpha-3}$, which cannot be made consistent
with~(\ref{magf.86}). Finally, when $\alpha,\beta<3$ and
$\alpha+\beta>3$ the integral ${\cal I}_{\alpha\beta}(0,\infty)$
converges (in the {\it light blue shaded} triangle in 
figure~\ref{convergence}) and can be evaluated~\cite{GradshteynRyzhik}
to yield ${\cal I}_{\alpha\beta}(0,\infty) = B(2-\alpha,\alpha+\beta-4)
-B(3-\beta,2-\alpha)-B(\alpha+\beta-4,3-\beta)$, where 
$B(\mu,\nu)=\Gamma(\mu)\Gamma(\nu)/\Gamma(\mu+\nu)$ denotes 
the beta function. Now since $|\beta- 3/2|\lsim 0.2$, Eq.~(\ref{magf.90}) 
can yield only a tiny amplification of the vacuum spectrum, such that
$\alpha\approx 1/2$, inconsistent with $\alpha+\beta>3$.

 To conclude, in this section we have reconsidered the coupling
of gauge fields to the scalar metric
perturbations~\cite{Maroto:2000zu} and showed that there is no
significant amplification of gauge fields in inflation. 
We first analysed the problem in the mean field approximation
(which overestimates the effect), and then we estimated the effect
of dissipation from inelastic scatterings by employing
a simple approximation for the spectra of $\Phi$ and $A_i$. 
The disagreement with Ref.~\cite{Maroto:2000zu} is probably
due to the use of the time dependent perturbation theory 
in~\cite{Maroto:2000zu}, which often leads to spurious
secular effects when applied in large time limit.

\section{Photon at preheating}
\label{Photon at preheating}

Since at the moment there is no satisfactory account of both the resonant
inflaton decay and plasma conductivity at preheating, in the present
work most of the times we make the conservative assumption that no
amplification occurs at preheating epoch. For an heuristic account
of this problem see for example
Ref.~\cite{DavisDimopoulosProkopecTornkvist-II},
and for a numerical work see~\cite{Bassett-Rajantie}.
Since explosive particle production is quite typical for many models 
of preheating, abundant production of charged particles occurs
quite generically immediately after inflation. The sudden matching 
approximation~\cite{Ratra:1992bn} may thus be appropriate, according to
which the conductivity becomes large quickly at the beginning
of radiation epoch such that the electric field generated in inflation 
decays, while the magnetic field gets frozen in. In this work we adopt
this simple approximation. For a recent reformulation of the matching
conditions appropriate for treatment of superhorizon charged scalar 
fluctuations created in inflation~\cite{Calzetta:1998ku} in terms of
more sophisticated field theoretical techniques 
see Ref.~\cite{Giovannini:2000dj}.

\section{Discussion and concluding remarks}
\label{Discussion and concluding remarks}

 We have considered various effects that break conformal invariance 
of gauge fields in inflation. We first analyse
the Drummond-Hathrell action in
section~\ref{Photon coupling to gravity with heavy fermions}
which describes a local one-loop modification to the gauge field 
action in presence of massive fermions, and find only a very modest
amplification of gauge fields. On the other hand in
section~\ref{Photon coupling to gravity with light fermions}
we consider the Dolgov trace anomaly of gauge fields
in presence of massless fermions and find a significant
gauge field amplification in inflation,
provided the number of light fermions is of the order a few hundred. 
We have then in section~\ref{Photon coupling to scalar field}
considered modified kinetic gauge terms by 
coupling to a scalar field and found that under quite 
reasonable conditions on the couplings one gets significant 
amplification of gauge fields in inflation, leading to potentially
observable magnetic fields on cosmological scales. 
In section~\ref{Photon coupling to pseudoscalar field} we consider 
coupling of a pseudoscalar field to the Chern-Pontryagin density, and
find only a mild amplification of the gauge field amplitude in inflation,
while the vacuum spectrum remains unchanged. We find in this case 
a nice example of birefringence induced by gravity, where one photon
polarization exhibits tachyonic growth in inflation, while the other,
which corresponds to a massive excitation, remains of a constant amplitude. 
Finally, in 
section~\ref{Photon coupling to scalar cosmological perturbations} 
we consider gauge field coupling to scalar metric 
perturbations~\cite{Maroto:2000zu} and find that the effect is by far 
too weak to produce observable magnetic fields.  

When this work was nearing completion, Ref.~\cite{Giovannini:2001xk}
appeared which overlaps in part with 
section~\ref{Photon coupling to scalar field}. 
The results of section~\ref{Photon coupling to scalar field} were presented
earlier at the workshop Beyond the Standard Model in Bad Honnef (March 2001).

\section*{Acknowledgements}

I wish to thank Anne Davis, Kostantinos Dimopoulos, and Ola T\"ornkvist for 
collaboration on related issues and for useful comments.
I thank Gert Aarts for helpful discussions.
I am indebted to Ola T\"ornkvist for discussions of
the gauge field amplification from the Dolgov anomaly analysed in 
section~\ref{Photon coupling to gravity with light fermions}.

\section*{Appendix A}

Here we present derivation of the photon field equation of motion starting
with the Drummond-Hathrell action~(\ref{magf.1}). 
In a conformally flat space-time the equation of motion can be recast as
\begin{eqnarray}
\eta^{\mu\rho}\partial_\mu F_{\rho\nu} 
 + \frac{\beta_e}{M^2}\partial_\mu\left[  
  b\eta^{\mu\rho}{\cal R}F_{\rho\nu}
   + \frac{c}{2} a^2\left({\cal R}^{\mu\rho} + {\cal R}^{\rho\mu}\right)
        F_{\rho\nu}
     + d a^4 \eta_{\gamma\nu}{\cal R}^{\mu\gamma\rho\sigma}F_{\rho\sigma}
  \right]
   = 0
 \, .
\label{magf.A1}
\end{eqnarray}
The metric of a conformally flat space-time can be in general written as
$g_{\mu\nu}=a^2(\eta)\eta_{\mu\nu}$, where 
$\eta_{\mu\nu}={\rm diag}(1,-1,-1,-1)$ is the Minkowski metric, 
and the scale factor $a=a(\tau)$ is a function of conformal time $\tau$. 
The nonvanishing components of the Riemann tensor have the following
simple form  
\begin{eqnarray}
{\cal R}_{0i0}^{\quad j} 
 &=& -  {\cal R}_{i00}^{\quad j}
 =  - \eta_i^{\;j}\left(
\frac{\partial_\tau ^2 a}{a}
 - \left(\frac{\partial_\tau a}{a}\right)^2 \right)
\nonumber\\
  {\cal R}_{i0j}^{\quad 0}
 &=& - {\cal R}_{0ij}^{\quad 0}
 =  - \eta_{ij}\left(
\frac{\partial_\tau ^2 a}{a}
 - \left(\frac{\partial_\tau a}{a}\right)^2 \right)
\nonumber\\
{\cal R}_{ijk}^{\quad l} 
 &=& \left(\eta_i^{\;l}\eta_{jk} - \eta_{ik}\eta_{j}^{\;l} \right)
      \left(\frac{\partial_\tau a}{a}\right)^2 
\label{magf.A2}
\end{eqnarray}
such that the curvature tensor components are
\begin{eqnarray}
{\cal R}_{00}
 &=& {\cal R}^i_{\;0i0}
 =  - 3 \left(\frac{\partial_\tau ^2 a}{a}
 - \left(\frac{\partial_\tau a}{a}\right)^2 \right)
\nonumber\\
{\cal R}_{ij}
 &=& {\cal R}^0_{\;i0j} + {\cal R}^l_{\;ilj}
 = - \eta_{ij}\left(
\frac{\partial_\tau ^2 a}{a}
 + \left(\frac{\partial_\tau a}{a}\right)^2 \right)\,,
\label{magf.A3}
\end{eqnarray}
and the Ricci curvature scalar reads
\begin{equation}
{\cal R} = - 6\,\frac{\partial_\tau ^2 a}{a^3} 
   = - 8\pi G \rho \left(1-3 w\right) ,
\label{magf.A4}
\end{equation}
where $w=p/\rho$ ($-1\leq w<1$).
To get the last equality we used Einstein's equation
$R_{\mu\nu} -{\cal R}g_{\mu\nu}/2 = 8\pi G T_{\mu\nu}$ 
which, in a conformal space time with the metric
$g_{\mu\nu} = a^2\eta_{\mu\nu}$, read
\begin{eqnarray}
H^2\equiv \left(\frac{\partial_\tau a}{a^2}\right)^2
  &=&  \frac{8\pi G}{3} \rho \qquad ({\rm Friedmann})
\nonumber\\
\frac{\partial_t^2 a}{a}
\equiv \frac{\partial_\tau^2 a}{a^3}
  - \left(\frac{\partial_\tau a}{a^2}\right)^2
  &=&  - \frac{4\pi G}{3} \left( \rho + 3 p\right) \,,
\label{magf.A5}
\end{eqnarray}
where we used the ideal fluid stress energy tensor
$T^\mu_\nu={\rm diag}(\rho,-p,-p,-p)$, where $\rho$ and $p$ denote the (total)
energy density and pressure, respectively. 
When $\rho + 3p < 0$ ($-1\leq w < -1/3$) the Universe undergoes an accelerating
expansion (inflation); the limiting case $w=-1$ corresponds to de Sitter
space-time, while matter and radiation era are obtained when
$w=0$ and $w=1/3$. More generally, $\rho\propto a^{-3(1+w)}$, such 
that upon integrating the Friedmann equation one obtains 
$a\propto \tau^{2/(1+3w)}$ (in de Sitter inflation $a= -1/H\tau$). 

Now making use of equations~(\ref{magf.A2} -- \ref{magf.A4}) 
and working in Coulomb gauge in which $A_0= 0 = \nabla \cdot \vec A$,
we simplify equation~(\ref{magf.A1}) to get 
\begin{eqnarray}
&&
\left(\partial_\tau^2 -\nabla^2\right)  A_i^{\,T}  
+ \frac{\beta_e}{M^2}\biggl[  
 (6b + c) \frac{\partial_\tau ^2 a}{a^3}
   + (c + 2d) \left(\frac{\partial_\tau a}{a^2}\right)^2 \;
  \biggr] \nabla^2 A_i^{\,T}
\nonumber\\
 &+& \frac{\beta_e}{M^2}\partial_\tau\biggl[  
 -(6b + 3c + 2d) \frac{\partial_\tau ^2 a}{a^3} \partial_\tau A_i^{\;T}
  + (3c + 2d)\left(\frac{\partial_\tau a}{a^2}\right)^2 \partial_\tau A_i^{\;T}
  \biggr]
   = 0
 \,,
\label{magf.A6}
\end{eqnarray}
where we used $F_{0i} =\partial_\tau A_i^{\;T}$ and 
$\partial_jF_{ji} =\nabla^2 A_i^{\;T}$, and 
$A_i^{\;T} = (\delta_{ij}-\partial_i\partial_j/\nabla^2)A_j$ denotes
the transverse photon field. 
Making use of Einstein's equations~(\ref{magf.5}), equation~(\ref{magf.A2})
can be recast as  
\begin{eqnarray}
&&
\left(\partial_\tau^2 -\nabla^2\right)  A_i^{\,T}  
+ \frac{\beta_e}{M^2 M_P^2}\biggl[  
 \frac{1}{6}(6b + 3c + 4d)\rho  -\frac{1}{2} (6b + c) p
  \biggr] \nabla^2 A_i^{\,T}
\nonumber\\
 &+& \frac{\beta_e}{M^2 M_P^2}\partial_\tau\left[  
\left(-\frac{1}{6}(6b - 3c - 2d)\rho  + \frac{1}{2}(6b + 3c + 2d)p \right)
   \partial_\tau A_i^{\;T}
  \right]
   = 0
 \,,
\label{magf.A7}
\end{eqnarray}
where $M_P\equiv (8\pi G)^{-1/2}=2.4\times 10^{18}$GeV is the reduced 
Planck mass. With $H^2 = \rho/3M_P^2$, $p=w\rho$, and 
$b=-5,c=26$ and $d=-2$, we finally arrive at  
\begin{equation}
\partial_\tau\left[\left(1 +  2\beta_e \frac{H^2}{M^2} \left(26 + 33w\right)
            \right)  \partial_\tau A_i^{\,T}\right]  
- \left(1 -  2\beta_e \frac{H^2}{M^2 } (10 + 3w)\right) \nabla^2 A_i^{\,T}  
      = 0  \,.\quad
\label{magf.A8}
\end{equation}
In section~\ref{Photon coupling to gravity with heavy fermions} we use
this equation to study the time evolution of the photon field in inflation. 

 In order to make a comparison with the photon coupling to 
a scalar field studied in Ref.~\cite{MazzitelliSpedalieri:1995}, 
first note that equation~(\ref{magf.A7}) yields    
\begin{eqnarray}
&&\partial_\tau\left[\left(1 +  \frac{15}{16}\beta_e \frac{H^2}{M_s^2} 
\left(1-6\xi\right)\left(1 - 3w\right)
            \right)  \partial_\tau A_i^{\,T}\right]  
\nonumber\\
&-& \left(1 + \frac{3}{16} \beta_e \frac{H^2}{M_s^2 } 
  \left[7-30\xi -(13-90\xi)w\right]
  \right) \nabla^2 A_i^{\,T}  
      = 0,
\label{magf.A9}
\end{eqnarray}
where we made use of the Schwinger-DeWitt
coefficients $b_s=(-5/16) + (15/8)\xi$, $c_s=1/4$ and $d_s=-3/8$,
and $M_s$ is the scalar field mass.
This case is quite different from the fermionic case studied 
in section~\ref{Photon coupling to gravity with heavy fermions} in that 
for $ \xi\leq 1/6$ the photon field is damped in inflation.
For a conformally coupled scalar field Eq.~(\ref{magf.A9}) reduces to 
\begin{equation}
   \partial_\tau^2 A_i^{\,T}
 - \left(1 + \frac{3}{8} \beta_e \frac{H^2}{M_s^2 } 
   \left(1 + w\right)
   \right) \nabla^2 A_i^{\,T}  
      = 0,\qquad   \xi= \frac 16
\label{magf.A10}
\end{equation}
and no significant amplification occurs in inflation. 
On the other hand when $\xi>1/6$ the photon coupled to a scalar field 
the amplitude grows in inflation. In this case the gauge field
amplification can be obtained by effecting the following replacement 
in Eq.~(\ref{magf.8}):
\begin{equation}
\iota_w \rightarrow \iota_{\xi w} 
 =  \frac{15}{16}\beta_e \left(6\xi-1\right)\left(1-3w\right)
  \frac{H^2}{M_s^2}, \qquad \xi> \frac 16\,.
\label{magf.A11}
\end{equation}
When both massive fermionic and scalar fields are present, one should 
add contributions from both fields. In de Sitter inflation
($w=-1$ and $H=$~const) Eq.~(\ref{magf.A9}) simplifies to 
\begin{equation}
\left(1 +  \frac{15}{4}\beta_e \frac{H^2}{M_s^2}\left(1-6\xi\right)
            \right)  \left(\partial_\tau^2 - \nabla^2\right) A_i^{\,T}  
      = 0,
\label{magf.A12}
\end{equation}
and thus conformal invariance is recovered, just like in the fermionic
case~(\ref{magf.4}).

\section*{Appendix B}

 We now outline computation of a volume averaged magnetic field strength
for a given gauge field spectrum of the form 
\begin{equation}
{\cal A}_{\vec k} = {\cal C}_0 V^{\frac 12} k^{-\alpha}  \cos (k\tau+\varphi_0)\,,
\label{magf.B1}
\end{equation}
where  $\alpha = \theta_w +1/2$ for the particular 
spectrum~(\ref{magf.21}) --~(\ref{magf.22}), and we displayed explicitly
the volume $V$ dependence, which was for simplicity ignored in the Wronskian 
normalization~(\ref{magf.13}). 
Following Ref.~\cite{DavisDimopoulosProkopecTornkvist} we {\it define} 
the volume averaged magnetic field correlated on a scale $\ell$ as~:  
\begin{eqnarray}
&&B_\ell^2 =
\langle B_i(\ell,\vec{x}\,) B_i(\ell,\vec{x}\,)\rangle -
\langle B_i(\ell,\vec{x}\,) B_i(\ell,\vec{x}\,)\rangle_{\rm vac}~,
\nonumber\\
&&B_i(\ell,\vec{x}\,) =\frac{3}{4\pi\ell^3}
\int_{|\vec{y}-\vec{x}|\leq\ell} d^3y B_i(\vec{y}\,)~,
\label{magf.B2}
\end{eqnarray}
where the average $\langle\cdot\rangle$ is
taken over Fock space as well as the position $\vec{x}$, and we subtracted
the (divergent) vacuum contribution $\langle\cdot\rangle_{\rm vac}$. 
The definition~(\ref{magf.B2}) corresponds to a sharp ball window function
$w(\vec x-\vec y,\ell)= 1$ for $|\vec x-\vec y\,|\leq 1$, and {\it zero}
otherwise. 

 Ignoring for the moment the vacuum contribution, and making use of 
the Fourier transform 
\begin{equation}
B_i(\vec y,\tau) =   \int {d^3k^{\,\prime}}{(2\pi)^3}
   e^{i\vec k^{\,\prime}\cdot\,\vec y} 
   \, i\epsilon_{ijl}k^{\,\prime}_j {\cal A}_{\,l} (\vec k^{\,\prime},\tau)
\,,\qquad  {\cal A}_{\,l} (\vec k^{\,\prime},\tau) 
  = \epsilon_{l}^T {\cal A}_{\vec k^{\,\prime}}(\tau)
\,,
\label{magf.B3}
\end{equation}
where $\epsilon_{l}^T$ denotes the unit transverse polarization 3-vector.
To evaluate the incurring integrals it is convenient to introduce the
following new variables~: $\vec r_1=\vec y - \vec x$, 
$\vec r_2 = \vec z - \vec x$, 
$\vec K = (\vec k^{\,\prime} -\vec k^{\,\prime\prime})/2$ and 
$\vec k = (\vec k^{\,\prime} -\vec k^{\,\prime\prime})/2$, such that 
after some algebra we obtain 
\begin{equation}
B_\ell^2  = \left(\frac{3}{4\pi\ell^2}\right)^2 \frac {1}{8V}
 \int_{k_0}^{\Lambda} \frac{dk}{k^2} 
 \langle {\cal A}_{\vec k}{\cal A}_{-\vec k}\rangle
  \left(2k\ell\cos 2k\ell - \sin 2k\ell\right)^2\,,
\label{magf.B4}
\end{equation}
where we introduced the infrared and ultraviolet cut-offs, $k_0$ and $\Lambda$,
respectively, which can be sent to infinity at the end of calculation.
Keeping finite cutoffs is useful for discussion of the divergent limits,
$\alpha_{\rm vac}=1/2$ (vacuum spectrum) and $\alpha_{\rm\,sc.inv.}=5/2$
(scale-invariant spectrum). Equation~(\ref{magf.B4}) can be further simplified 
\begin{equation}
B_\ell^2  = \frac{9\times 2^{2\alpha-7}}{\pi^2}\; {\cal C}_0^2\,\iota_\alpha\,,
\qquad
\iota_\alpha (x_0,x_1) = \int_{x_0}^{x_1} \frac{dx}{x^{2(\alpha-1)}} j_1^2(x)
\,,
\label{magf.B5}
\end{equation}
where $x_0=2k_0\ell$, $x_1=2\Lambda\ell$, 
$j_1=dj_0/dx=(\cos x - \sin x/x)/x$ denotes the spherical Bessel function,
and we used
the long time-average 
$\langle \cos^2 (k\tau+\varphi_0)\rangle_\tau =1/2$. When $1/2 <\alpha < 5/2$
the integral $\iota_\alpha(x_0\rightarrow 0,x_1\rightarrow\infty)$ converges
and can be evaluated exactly ({\it cf.} Eqs. 6.576.2 and 9.122.1
in Ref.~\cite{GradshteynRyzhik}). The result is
\begin{equation}
\iota_\alpha (0,\infty) = \frac{\sqrt{\pi}}{4}
  \frac{\Gamma\left(\frac 52 -\alpha\right)\Gamma\left(\alpha-\frac 12\right)}
   {\Gamma\left(\alpha\right)\Gamma\left(\alpha + \frac 32\right)}\,,
\qquad 
\frac 12 <\alpha < \frac 52
\,.
\label{magf.B6}
\end{equation}
Putting everything together, we obtain 
\begin{equation}
 B_\ell = b_\alpha\, {\cal C}_0\, \ell ^{\,\alpha-\frac 52}
\,,\qquad
 b_\alpha^2 = \frac{9\times 2^{2\alpha-9}}{\pi^{3/2}}
\frac{\Gamma\left(\frac 52 -\alpha\right)\Gamma\left(\alpha-\frac 12\right)}
   {\Gamma\left(\alpha\right)\Gamma\left(\alpha + \frac 32\right)}
\,,
\qquad 
\frac 12 <\alpha < \frac 52
\,,
\label{magf.B7}
\end{equation}
such that $b_1= \sqrt{3}/(8\sqrt{\pi})$ for the thermal spectrum and 
$b_\frac 32 = 3/8\pi$ for the spectrum
$B_\ell\propto \ell^{-1}$.

 For the vacuum spectrum ($\alpha=1/2$) the integral $\iota_\frac 12$
diverges in the UV. Upon introducing an UV cut-off,
the leading-log contribution to the integral equals  
$\iota_\frac 12 = (1/2)\ln 2\Lambda\ell$, and hence the spectrum reads
\begin{equation}
 B_{\ell}^{\rm\,vac} = \frac{3}{8\pi\sqrt{2}}\; {\cal C}_0\, 
 (\ln 2\Lambda\ell)^\frac 12\, \ell ^{\,-2}\,,
\qquad \alpha = \frac 12
\,.
\label{magf.B8}
\end{equation}
The question is of course what is the natural value for the UV cut-off.
At a first sight it may seem reasonable to choose the Hubble scale
$\Lambda\sim H$, or the Planck scale $\Lambda\sim M_P$. 
We believe however that it is more natural
to take $\Lambda\sim e/2\ell$, implying that measurements at some scale 
$\ell$ are mostly sensitive to the field excitations at that scale.
With this equation~(\ref{magf.B8}) becomes cut-off independent~:
\begin{equation}
 B_{\ell}^{\rm\,vac} \sim  \frac{3}{8\pi\sqrt{2}}\; {\cal C}_0\, 
  \, \ell ^{\,-2}\,,
\qquad \alpha = \frac 12
\,.
\label{magf.B9}
\end{equation}
%

 For the scale invariant spectrum $\alpha = 5/2$, the integral $\iota_\alpha$ 
becomes infrared dominated, suggesting an infrared cut-off, $k_0$. 
To leading-log accuracy $\iota_\alpha(x_0,\infty) = -(1/9)\ln x_0$ 
($x_0=2k_0\ell$), so that 
\begin{equation}
 B_{\ell}^{\rm\,sc.inv.} =  \frac{1}{2\pi}\; {\cal C}_0\, 
  \left(-\ln 2k_0\ell\right)^{\frac 12}\,,
\qquad \alpha = \frac 52
\,.
\label{magf.B10}
\end{equation}
In this case the natural cut-off is is the lowest momentum that gets
amplified in inflation, implying that information about the beginning
of inflation is present in the spectrum.
This should not be taken very seriously however, because in practice
it may be very difficult to distinguish a scale invariant spectrum 
from an almost scale invariant spectrum with $\alpha\lsim 5/2$ in 
which the dependence on the initial conditions in inflation drops out.
For this reason, when plotting the scale-invariant spectrum we take 
$k_0 = e/2\ell$, for which equation~(\ref{magf.B10}) becomes truly scale
invariant, $B_{\ell}^{\rm\,sc.inv.} \sim {\cal C}_0/2\pi$.

\section*{Appendix C}

Here we consider two simple inflationary models:
chaotic inflation and extended inflation. 

\subsubsection*{C.1 Chaotic inflation with $V(\phi)=m^2\phi^2/2$}
\label{Chaotic inflation}

 For simplicity we assume that $\phi$ is the inflaton with the potential 
$V(\phi)=m^2\phi^2/2$~\cite{Linde:1983gd}, where the cosmic microwave 
background radiation (CMBR) measurements constrain 
$m\approx 1.8\times 10^{13}$~GeV. The considerations in this 
section can be easily generalized to any power-law potential 
$V_n(\phi)=(\lambda_n/n!)(\phi/M_P)^n$, 
with $\lambda_4\approx 5 \times 10^{-13}$, {\it etc}. 
The equation of motion for the inflaton $\phi$ reads
\begin{equation}
\frac{d^2\phi}{dt^2}  + 3H\frac{d\phi}{dt} + m^2\phi 
      = 0,
\label{magf.c1}
\end{equation}
where $H^2=\rho/3M_P^2$, and $\rho$ is the total energy density. 
Equation~(\ref{magf.c1}) can be easily solved in the slow-roll approximation
\footnote{For more formal treatment of the slow-roll
conditions see for example~\cite{Lyth:1999xn}.}
in which ${d^2\phi}/{dt^2}\ll 3H {d\phi}/{dt}, \;\partial_\phi V$ and
$({d\phi}/{dt})^2\ll 2V$ such that 
\begin{equation}
 H\equiv  \frac 1a \frac {da}{dt}=\frac {m\phi}{\sqrt{6}M_P}.
\label{magf.c2}
\end{equation}
The result is 
\begin{equation}
  \phi  = \phi_0 - \sqrt {\frac 23} M_P\, mt
,
\label{magf.c3}
\end{equation}
where $\phi_0$ can be related to the number of $e$-folds before 
the end of inflation, $\phi_0^2(N)\approx 4\sqrt{6}M_P^2N$.
Integrating~(\ref{magf.c2}) once we then get for the scale factor 
\begin{eqnarray}
\ln \frac {a}{a_0} &=& \frac{m\phi_0t}{\sqrt{6}M_P} - \frac{m^2t^2}{6}
\nonumber\\
&\approx& H t
.
\label{magf.c4}
\end{eqnarray}
%
The time dependence of $H$ becomes important only at late times in
inflation.\footnote{An alternative definition is the following 
time averaged expansion rate: 
$\bar H= H_0-m^2t^2/6 \approx H_0 - (m^2/6H_0)\ln (-H_0\tau)$, 
where $H_0={m\phi}/{\sqrt{6}M_P}$. This definition gives the exact
result for the scale factor~(\ref{magf.c4}) when expressed in terms
of cosmic time $t$.} 
This then immediately implies 
\begin{equation}
a \approx   -\frac{1}{H\tau}
\label{magf.c5}
\end{equation}
and 
\begin{eqnarray}
\frac{d\phi}{d\tau}  = a\frac{d\phi}{dt} &\approx& 
        \frac{2M_P^2}{\phi}\frac {1}{\tau}
\label{magf.c6}
\end{eqnarray}
where $\phi^2(\tau) \approx 4\sqrt{6}M_P^2 N$.

\subsection*{C.2 Extended inflation}
\label{Extended inflation}

Extended inflation~\cite{La:1989st} is realised by an
exponential potential $V(\phi)=$  $M_P^4 e^{-\lambda'\phi/M_P}$,
and it corresponds to the Jordan-Brans-Dicke theory of gravity
rewritten in the Jordan frame. For an alternative formulation 
see~\cite{JoyceProkopec:1998}. In its original version the model
has limitations related to the graceful exit problem and the CMBR 
constraints. The spectral slope $n$ of scalar cosmological 
perturbations $n-1=-{\lambda'}^{\,2}$, and the fraction of 
perturbations in gravitational radiation $r=5{\lambda'}^{\,2}$ are
constrained to be much smaller than unity~\cite{Lyth:1999xn}. Here we
assume that the former problem is solved by matching at the end inflation 
onto a power law potential, while the latter can be resolved 
by an appropriate choice of the coupling constant $\lambda'$. We discuss
extended inflation primarily for its simplicity, and 
focus on the attractor solution~\cite{RatraPeebles:1988,JoyceProkopec:1998}
 because it corresponds to more generic
initial conditions than the slow-roll case. 

The inflaton equation of motion reads
\begin{equation}
  \frac{d^2\phi}{dt^2}  + 3H\frac{d\phi}{dt} 
      - \lambda' \frac{V_0}{M_P} e^{-\lambda'\phi/M_P}
      = 0,
\label{magf.c7}
\end{equation}
where 
\begin{equation}
H^2 = \frac{1}{2}\left(\frac{d\phi}{dt}\right)^2 + V(\phi)
.
\label{magf.c8}
\end{equation}
It is quite easy to show that equations~(\ref{magf.c7}-\ref{magf.c8}) are 
solved by the following attractor solution:  

\begin{eqnarray}
\frac{d\phi}{dt} &=& \frac{2M_P}{\lambda' t}
\nonumber\\
  \phi &=& \frac{2M_P}{\lambda'}\left(\ln\frac{\sqrt{V_0}\;t}{M_P}
        -  \frac{1}{2}\ln\left[\frac{2}{{\lambda'}^{\,2}}
           \left(\frac{6}{{\lambda'}^{\,2}}-1\right)\right]\right)
,
\label{magf.c9}
\end{eqnarray}
such that 
\begin{equation}
 V  = \frac{2}{{\lambda'}^{\,2}}
  \left(\frac{6}{{\lambda'}^{\,2}}-1\right) \frac{M_P^2}{t^2}
,
\label{magf.c10}
\end{equation}
and hence 
\begin{equation}
 H \equiv \frac{1}{a}\frac{da}{dt} = \frac{2}{{\lambda'}^{2} t}
\,.
\label{magf.c11}
\end{equation}
Integrating this once gives 
\begin{equation}
 a  = \left(M_0 t\right)^{{2}/{{\lambda'}^{2}}}
,
\label{magf.c12}
\end{equation}
where $M_0 < M_P $ is a mass scale that characterises the beginning of 
inflation, $t_0\sim 1/M_0$.
This immediately implies that $\lambda' <\sqrt{2}$ 
is required to get a superluminal expansion. On the other hand 
to get a nearly scale-invariant spectrum of perturbations, $\lambda'\ll 1$. 
Since inflation is assumed to terminate at $\phi\approx 0$,
in order to get more than 60 $e$-folds (required to solve the standard 
cosmological problems), the potential $V_0$ should be sufficiently small, 
such that at the beginning of inflation we have $V(\phi_0)\ll M_P^4$. 
Now making use of $ad\tau =dt$ we can reexpress (\ref{magf.c12}) as 
\begin{equation}
 a  = \left(- \left(\frac{2}{{\lambda'}^{2}}-1\right) M_0 \tau\right)
       ^{-\frac{2}{2-{\lambda'}^{2}}}
,
\label{magf.c13}
\end{equation} 
so that 
\begin{equation}
\frac{d\phi}{d\tau}  = a \frac {d\phi}{dt}
 = - \frac{2M_P}{\lambda'\left(\frac{2}{{\lambda'}^{2}} -1\right) }
     \frac {1}{\tau}
.
\label{magf.c14}
\end{equation}
Finally, equation~(\ref{magf.c9}) can be expressed as a function of 
conformal time as follows: 
\begin{eqnarray}
\phi &=& \phi_0
 - \frac{2\lambda'}{2-{\lambda'}^2} M_P
  \ln\left[ -\left(\frac{2}{{\lambda'}^2}-1\right)M_0\tau\right]
\nonumber\\
 \phi_0 &=&
 -\frac{M_P}{\lambda'}\ln\left[\frac{2}{{\lambda'}^2}
    \left(\frac{6}{{\lambda'}^2}-1\right)\frac{M_P^2M_0^2}{V_0}\right]
\label{magf.c15}
\end{eqnarray}
Equations~(\ref{magf.c6}) and~(\ref{magf.c14}) are used in 
section~\ref{Photon coupling to scalar and pseudoscalar fields}
for computation of magnetic field amplification in inflation.

%
%

\nc{\ap}[3]    {{\it Ann.\ Phys.\ }{{\bf #1}, {#2} {(#3)}}}
\nc{\ibid}[3]  {{\it ibid.\ }{{\bf #1}, {#2} {(#3)}}}
\nc{\jmp}[3]   {{\it J.\ Math.\ Phys.\ }{{\bf #1}, {#2} {(#3)}}}
\nc{\np}[3]    {{\it Nucl.\ Phys.\ }{{\bf #1}, {#2} {(#3)}}}
\nc{\pl}[3]    {{\it Phys.\ Lett.\ }{{\bf #1}, {#2} {(#3)}}}
\nc{\pr}[3]    {{\it Phys.\ Rev.\ }{{\bf #1}, {#2} {(#3)}}}
\nc{\prd}[3]    {{\it Phys.\ Rev.\ }D {{\bf #1}, {#2} {(#3)}}}
\nc{\prep}[3]  {{\it Phys.\ Rep.\ }{{\bf #1}, {#2} {#3)}}}
\nc{\prl}[3]   {{\it Phys.\ Rev.\ Lett.\ }{{\bf #1}, {#2} {(#3)}}}
\nc{\spjetp}[3]{{\it Sov.\ Phys.\ JETP }{{\bf #1}, {#2} {(#3)}}}
\nc{\zetp}[3]  {{\it Zh.\ Eksp.\ Teor.\ Fiz.\ }{{\bf #1}, {#2} {(#3)}}}

\newcommand{\jf}{\it}
\newcommand{\jt}{\/}
\newcommand{\VPY}[3]{{\bf #1}, #2 (#3)}
\newcommand{\ispace}{\thinspace}
\let\U=\.
\def\.{.\nobreak\ispace\ignorespaces}
\newcommand{\IJMODPHYS}[3]{{\jf Int. J. Mod. Phys.\jt} \VPY{#1}{#2}{#3}}
\newcommand{\JETP}[3]{{\jf JETP Lett.\jt} \VPY{#1}{#2}{#3}}
\newcommand{\MPL}[3]{{\jf Mod. Phys. Lett.\jt} \VPY{#1}{#2}{#3}}
\newcommand{\NC}[3]{{\jf Nuovo Cim.\jt} \VPY{#1}{#2}{#3}}
\newcommand{\Nature}[3]{{\jf Nature} \VPY{#1}{#2}{#3}}
\newcommand{\NP}[3]{{\jf Nucl. Phys.\jt} \VPY{#1}{#2}{#3}}
\newcommand{\PHYREP}[3]{{\jf Phys. Rept.\jt} \VPY{#1}{#2}{#3}}
\newcommand{\PL}[3]{{\jf Phys. Lett.\jt} \VPY{#1}{#2}{#3}}
\newcommand{\PR}[3]{{\jf Phys. Rev.\jt} \VPY{#1}{#2}{#3}}
\newcommand{\PRD}[3]{{\jf Phys. Rev. D\jt} \VPY{#1}{#2}{#3}}
\newcommand{\PRL}[3]{{\jf Phys. Rev. Lett.\jt} \VPY{#1}{#2}{#3}}
\newcommand{\PTRSLA}[3]{{\jf Phil. Trans. R. Soc. Lond.\jt\ A}
\VPY{#1}{#2}{#3}}
\newcommand{\ZhETF}[3]{{\jf Zh. Eksp. Teor. Fiz.\jt} \VPY{#1}{#2}{#3}}

\end{document}